%% file: main.tex
\documentclass[acmsmall,screen]{acmart}
\AtBeginDocument{%
  }
\usepackage[normalem]{ulem}

\setcopyright{acmlicensed}
\copyrightyear{2018}
\acmYear{2018}
\acmDOI{XXXXXXX.XXXXXXX}

\acmJournal{JACM}
\acmVolume{37}
\acmNumber{4}
\acmArticle{111}
\acmMonth{8}

\input{package}

\input{macro}

\begin{document}

\title{PRISM: \underline{P}hotonics-Info\underline{r}med \underline{I}nver\underline{s}e Lithography for \\ \underline{M}anufacturable Inverse-Designed Photonic Integrated Circuits}

\author{Hongjian Zhou}
\email{hzhou144@asu.edu}
\affiliation{%
  \institution{Arizona State University}
  \city{Tempe}
  \state{Arizona}
  \country{USA}
}

\author{Haoyu Yang}
\affiliation{%
  \institution{NVIDIA Corp.}
  \city{Austin}
  \state{Texas}
  \country{USA}
  }
\email{haoyuy@nvidia.com}

\author{Nicholas Gangi}
\email{gangin2@rpi.edu}
\affiliation{%
  \institution{Rensselaer Polytechnic Institute}
  \city{Troy}
  \state{New York}
  \country{USA}
}

\author{Tianle Xu}
\email{Xut6@rpi.edu}
\affiliation{%
  \institution{Rensselaer Polytechnic Institute}
  \city{Troy}
  \state{New York}
  \country{USA}
}

\author{Rena Huang}
\email{HUANGZ3@rpi.edu }
\affiliation{%
  \institution{Rensselaer Polytechnic Institute}
  \city{Troy}
  \state{New York}
  \country{USA}
}

\author{Jiaqi Gu}
\authornote{Corresponding author.}
\email{jiaqigu@asu.edu}
\affiliation{%
  \institution{Arizona State University}
  \city{Tempe}
  \state{Arizona}
  \country{USA}
}

\renewcommand{\shortauthors}{Zhou et al.}

\input{doc/0_abstract}

\setcopyright{none}
\renewcommand\footnotetextcopyrightpermission[1]{}
\settopmatter{printacmref=false}

\maketitle
\thispagestyle{plain}
\pagestyle{plain}

\input{doc/1_intro}

\input{doc/2_prelim}
\input{doc/3_method}

\input{doc/4_result}

\input{doc/5_conclu}
\input{doc/6_acknow}


\input{main.bbl}
\end{document}

%% file: package.tex
\usepackage[utf8]{inputenc} %
\usepackage[T1]{fontenc}    %

\usepackage{pifont}
\usepackage{xcolor}         %
\usepackage{booktabs}
\usepackage{hyperref}
\hypersetup{colorlinks,
    linkcolor=red,citecolor=citecolor,urlcolor=blue}
\usepackage{algorithm}
\usepackage{graphicx}
\usepackage{threeparttable}
\usepackage{multirow}
\usepackage{amsmath}
\usepackage{bm}
\usepackage{bbm}
\usepackage{amsfonts}
\usepackage{amsthm}
\usepackage[noend]{algpseudocode}
\usepackage{enumitem} %
\usepackage[subrefformat=parens,labelformat=parens]{subfig}
\usepackage{colortbl}
\usepackage[normalem]{ulem}
\usepackage{soul}
\usepackage{graphicx} %
\usepackage{booktabs}
\usepackage{algpseudocode}
\usepackage{subfig}
\algtext*{EndIf}

\usepackage{wrapfig}
\usepackage{algorithmicx}

\usepackage{xspace}

\algnewcommand{\REQUIRE}{\item[\algorithmicrequire]}
\algnewcommand{\ENSURE}{\item[\algorithmicensure]}

\algnewcommand{\RETURN}{\State \algorithmicreturn }

\algnewcommand{\algorithmicoptional}{\textbf{Optional:}}
\algnewcommand{\OPTIONAL}{\item[\algorithmicoptional]}

\algrenewcommand\algorithmiccomment[1]{\hfill{\footnotesize$\triangleright$ #1}}

%% file: macro.tex
\definecolor{citecolor}{RGB}{34,139,34}
\definecolor{mydarkblue}{rgb}{0,0.08,1}
\definecolor{mydarkgreen}{rgb}{0.02,0.6,0.02}
\definecolor{mydarkred}{rgb}{0.8,0.02,0.02}
\definecolor{mydarkorange}{rgb}{0.40,0.2,0.02}
\definecolor{mypurple}{RGB}{111,0,255}
\definecolor{myred}{rgb}{1.0,0.0,0.0}
\definecolor{mygold}{rgb}{0.75,0.6,0.12}
\definecolor{myblue}{rgb}{0,0.2,0.8}
\definecolor{mydarkgray}{rgb}{0.,0.2,0.2}

\definecolor{lightred}{RGB}{255,235,235}
\definecolor{lightgreen}{RGB}{235,255,235}
\definecolor{lightblue}{RGB}{235,235,255}
\definecolor{lightcyan}{RGB}{235,255,255}
\definecolor{lightmagenta}{RGB}{255,235,255}
\definecolor{lightyellow}{RGB}{255,255,235}

\definecolor{qxkcolor}{RGB}{215,235,255}
\definecolor{softmaxcolor}{RGB}{230,235,255}
\definecolor{probxvcolor}{RGB}{255,255,235}

\definecolor{topkcolor}{RGB}{255,235,235}
\definecolor{zecolor}{RGB}{255,255,235}
\definecolor{dynacolor}{RGB}{235,255,255}

\definecolor{reviewcolor}{RGB}{0,0,200}

\theoremstyle{plain}

\theoremstyle{definition}

\newcommand{\squishlist}{
 \begin{list}{$\bullet$}
  { \setlength{\itemsep}{0pt}
     \setlength{\parsep}{3pt}
     \setlength{\topsep}{3pt}
     \setlength{\partopsep}{0pt}
     \setlength{\leftmargin}{1.5em}
     \setlength{\labelwidth}{1em}
     \setlength{\labelsep}{0.5em} } }

\newcommand{\squishend}{
  \end{list}  }

\newcommand{\name}{\texttt{PRISM}\xspace}

%% file: doc/0_abstract.tex
\begin{abstract}
Photonic integrated circuits (PICs) are emerging as a key hardware platform for high-bandwidth, energy-efficient optical interconnects and for computing and sensing workloads increasingly relevant to large-scale AI systems.
Recent advances in photonic inverse design have demonstrated the ability to automatically synthesize compact, high-performance photonic components that surpass conventional, hand-designed structures, offering a promising path toward scalable and functionality-rich photonic hardware.
However, the practical deployment of inverse-designed PICs is bottlenecked by manufacturability: their irregular, subwavelength geometries are highly sensitive to fabrication variations, leading to large performance degradation, low yield, and a persistent gap between simulated optimality and fabricated performance.
Unlike electronics, where foundries provide mature design-for-manufacturing (DFM) pipelines that correct masks and shield designers from process details, photonics lacks a systematic, flexible mask optimization flow. 
Fabrication deviations in photonic components cause large optical response drift and compounding error in cascaded circuits, while calibrating fabrication models remains costly and expertise-heavy, often requiring repeated fabrication cycles that are inaccessible to most designers.
To bridge this gap, we introduce \name, a photonics-informed inverse lithography workflow that makes photonic mask optimization data-efficient, reliable, and optics-informed.
\name (i) synthesizes compact, informative calibration patterns to minimize required fabrication data, (ii) trains a physics-grounded differentiable fabrication digital twin, enabling gradient-based optimization, and (iii) performs photonics-informed inverse mask optimization that prioritizes performance-critical features beyond geometry matching.
Across multiple inverse-designed components with both electron-beam lithography and deep ultra-violet photolithography processes, \name significantly boosts post-fabrication performance and yield while reducing calibration area and turnaround time, enabling and democratizing manufacturable and high-yield inverse-designed photonic hardware at scale.
\end{abstract}


%% file: doc/1_intro.tex
\section{Introduction}
\label{sec:Introduction}
Photonic integrated circuits (PICs) are emerging as a foundational hardware platform for high-bandwidth communication~\cite{NP_CICC2024_Wang, NP_DATE2020_Thonnart}, optical interconnects~\cite{NP_CICC2024_Wang,NP_DATE2020_Thonnart,NP_TCAD2013_Ye,NP_ICCAD2022_Taheri,10323627}, and speed-of-light computing~\cite{zhou2022photonic, NP_NATURE2017_Shen, NP_Nature2025_Hua, NP_Nature2025_Ahmed}, with growing adoption in various key use domains, e.g., artificial intelligence (AI) systems, sensing, quantum, scientific discovery, etc.
Yet despite this promise, the practical design and deployment of \textbf{complex PICs at scale remain constrained by a manufacturing reality}: 
\emph{what is fabricated on wafer can deviate substantially from the intended design}.
For subwavelength photonic structures, e.g., gratings, inverse-designed patterns, these fine-grained structures often suffer from huge geometric deviations leading to large performance degradation and malfunction, as illustrated in Fig.~\ref{fig:Motivation_FabError}.
This becomes especially problematic in cascaded circuits, where small component-level errors can accumulate into system-level failure
and make inverse-designed layouts \textbf{difficult to manufacture reliably at scale}~\cite{ma2025boson, ONN_DAC2026_Zhou_SS}.

In production-oriented deep ultraviolet (DUV) fabrication processes in photonic foundries, e.g., 193 nm technology node, inverse-designed patterns can, in general, be considered \emph{infeasible to manufacture} due to severe distortions (e.g., rounding, blurring, bridging) that lead to \textbf{near-zero yield}.
Electron-beam lithography (EBL) can provide higher patterning resolution for laboratory demonstrations, but it remains costly and slow for high-volume manufacturing and still exhibits non-ideal distortion. 
As illustrated in Fig.~\ref{fig:Motivation_Yield}, DUV fabrication induces substantially larger figure of merit (FoM) degradation and a heavier worst-case tail than EBL across representative photonic devices, underscoring the manufacturability gap for inverse-designed layouts under DUV.
Figure~\ref{fig:Motivation_Variation} further shows that the wavelength division multiplexing transmission is jointly sensitive to lithography defocus and resist threshold, exhibiting a narrow high-performance process window.
Today, designers often cope through \textbf{expensive trial-and-error}: imposing conservative design rules, sacrificing geometric degrees of freedom with smoothing, running extensive parameter sweeps to search for a feasible design, and iteratively re-taping-out designs based on sparse fabrication feedback. 
This workflow is slow (year-long), high-cost, expertise-heavy, and fundamentally at odds with the promise of inverse design: \emph{it either sacrifices performance or fails to achieve robust yield}.

A \underline{natural question} is \emph{why photonics cannot simply adopt the mature design-for-manufacturing (DFM)~\cite{raghvendra2005dfm} stack from electronic design automation (EDA)}~\cite{huang2021machine}.
In electronics, foundries routinely apply optical proximity correction (OPC)~\cite{spence2005full} and inverse lithography (ILT)~\cite{pang2021inverse} to optimize masks and compensate for process effects.
In photonics, \textbf{two obstacles break this direct transfer}.
\ding{202}~Photonic hardware performance relies on complicated light propagation behavior; mask correction objectives based purely on geometric similarity are not sufficient, i.e., \textbf{“close geometry” does not imply “close function.”}
Optical response arises from nonlocal electromagnetic interactions (scattering, interference, resonance, etc) and is often dominated by a subset of sensitivity-critical features, so errors in a small region can dominate performance even when global pixel-wise error is small, illustrated in Fig.~\ref{fig:Motivation_Sensitive}.
Moreover, \emph{systematic} geometric biases, such as globally biased over-/under-etch, can be especially detrimental because they coherently shift effective widths, duty cycles, and phase accumulation across the device, leading to pronounced performance degradation even with a lower $L_2$ fabrication error, as shown in Fig.~\ref{fig:Motivation_L2}.
\ding{203}~\textbf{Accurate lithography modeling is expensive and process-specific}: foundry details are often proprietary, and purely data-driven fabrication surrogates can require large calibration datasets. When trained with limited data, black-box neural models may produce unreliable gradients, undermining gradient-based inverse lithography optimization where stability depends on smoothness, generalization, and physically plausible sensitivities, as shown in Fig.~\ref{fig:Motivation_GradErr}.
These gaps motivate a new approach: a \textbf{data-efficient, physically grounded, and photonics-informed inverse lithography workflow} that closes the gap between design and manufacturing by optimizing masks for photonic function fidelity beyond pattern matching, enabling high-yield fabrication capability of inverse-designed PICs for foundries while democratizing model calibration and mask optimization through low-cost, easily adaptable customization.
\input{figtex/fig_motivation}

In this work, we introduce \name, a photonics-informed neural inverse lithography framework that turns inverse-designed PIC layouts into high-yield, manufacturable hardware.
\name workflow unified three synergistic innovations:
(1) \textbf{data-efficient, identifiability-driven calibration pattern synthesis} strategy that generates compact, informative fabrication calibration patterns for training data acquisition.
(2) \textbf{a physics-grounded neural fabrication model} with superior generalization and provides stable gradients for mask optimization; and
(3) \textbf{photonics-informed ILT} that incorporates optical response as guidance to preserve performance-critical features beyond geometry matching.
Based on the approaches in \name, we make the following contributions:
\vspace{-3pt}
\squishlist
    {\item \textbf{Identifiability-driven Calibration Pattern Synthesis} -- We constructs a compact calibration reticle by combining (i) interpretable rectilinear/curvilinear regular motifs as controlled probes, (ii) randomized curvilinear patterns with controlled spectra for broad frequency/proximity coverage, and (iii) inverse-designed device crops for distribution matching, enabling accurate digital-twin calibration under a strict SEM/area budget.}
    {\item \textbf{Physics-grounded Differentiable Fabrication Modeling} -- We train a differentiable fabrication digital twin model that generalizes under limited data and provides stable gradients for mask optimization across patterns.}
    {\item \textbf{Photonics-informed Inverse Lithography} -- We formulate ILT as a function-preserving mask optimization problem and introduce EM-field-aware objectives to preserve photonic function beyond geometry fidelity.}
    {\item Across multiple inverse-designed components in EBL and 193 nm DUV photolithography, \name significantly boosts post-fabrication device performance and yield with substantially lower calibration cost, unlocking manufacturable inverse-designed photonic hardware at scale.}
\squishend

\input{figtex/fig_motivation2}

%% file: figtex/fig_motivation.tex
\begin{figure*}
    \centering
    \subfloat[]{\includegraphics[width=0.32\linewidth]{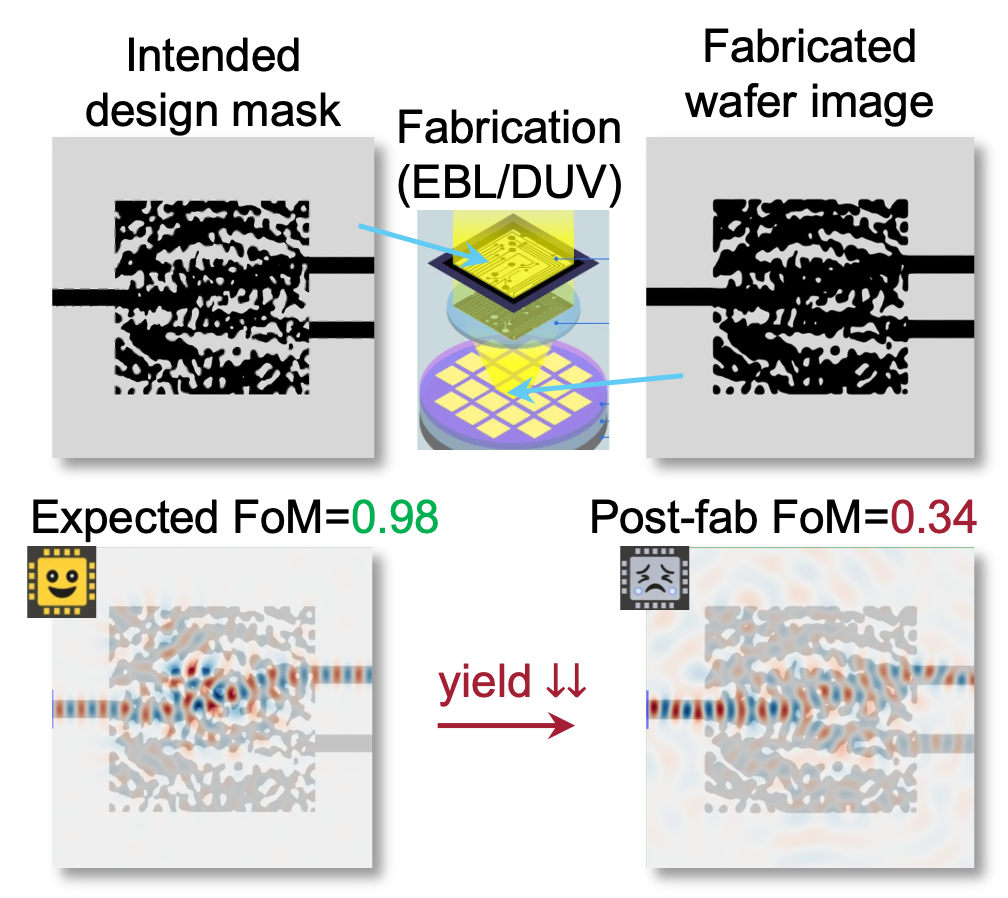}
    \label{fig:Motivation_FabError}
    }
    \subfloat[]{\includegraphics[width=0.335\linewidth]{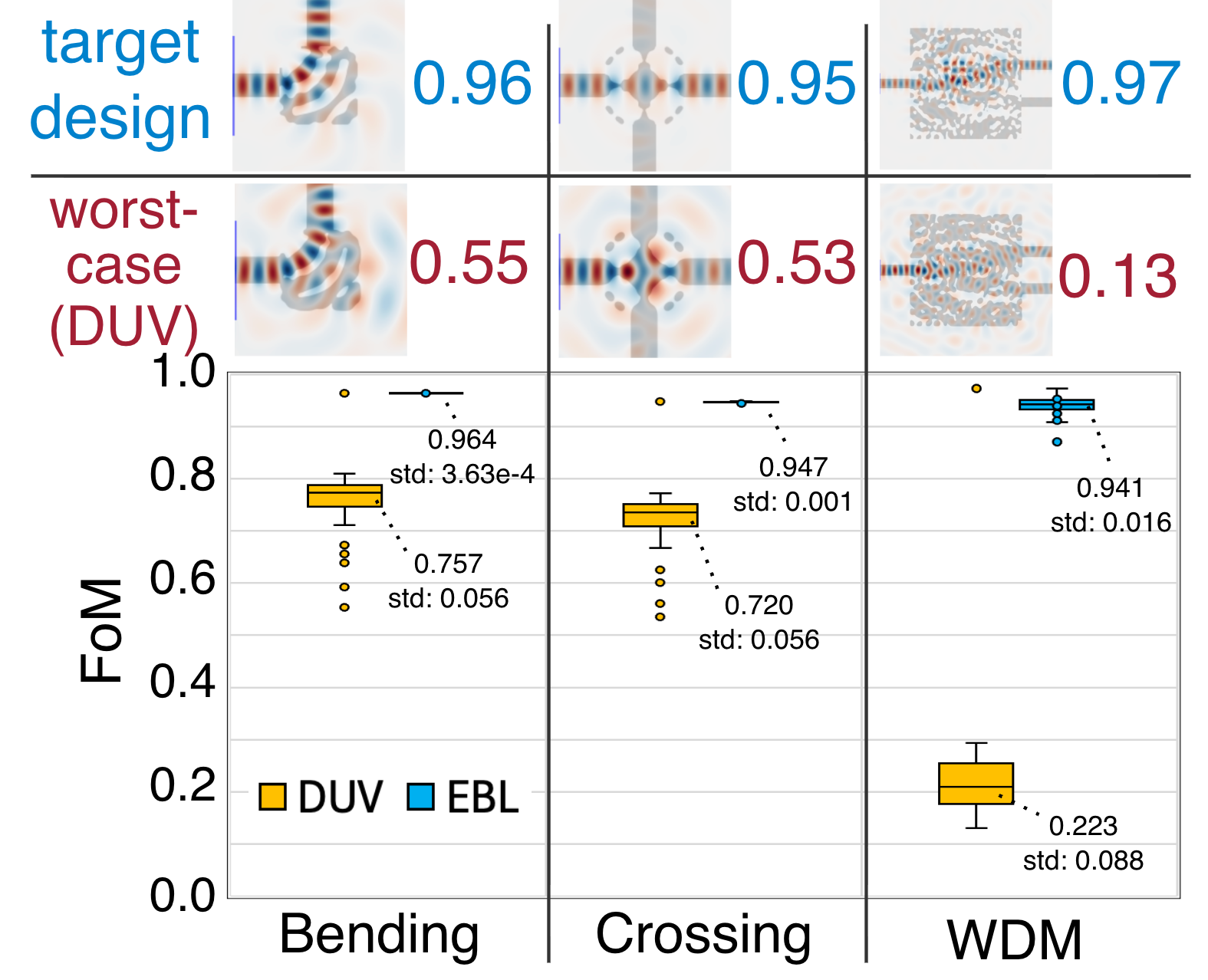}
    \label{fig:Motivation_Yield}
    }
    \subfloat[]{\includegraphics[width=0.32\linewidth]{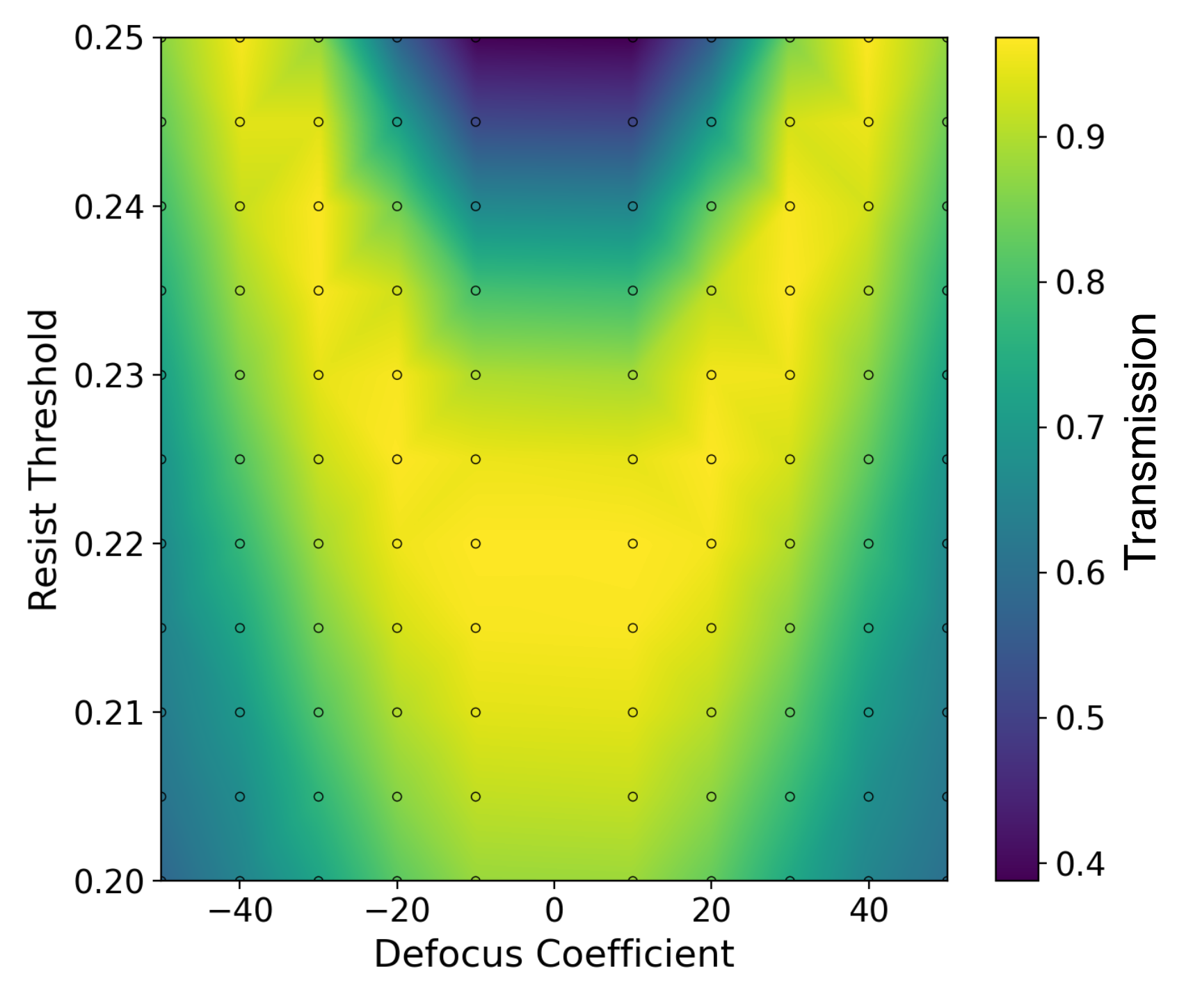}
    \label{fig:Motivation_Variation}
    }
    \vspace{-5pt}
    \caption{Comprehensive motivation. 
    (a) Fabrication error causes severe performance and yield degradation on inverse-designed photonic components.
    (b) DUV fabrication is particularly challenging for inverse-designed photonic devices, often causing substantially larger performance degradation and lower yield than EBL. Mean and std. of FoMs are marked.
    (c)Sensitivity map of device transmission under coupled process variations. The heatmap shows the transmission as a function of defocus coefficient and resist threshold, highlighting their joint impact on the post-fabrication response. 
    }
    \vspace{-10pt}
    \label{fig:Motivation}
\end{figure*}

%% file: figtex/fig_motivation2.tex
\begin{figure*}
    \centering
    \subfloat[]{\includegraphics[width=0.3\linewidth]{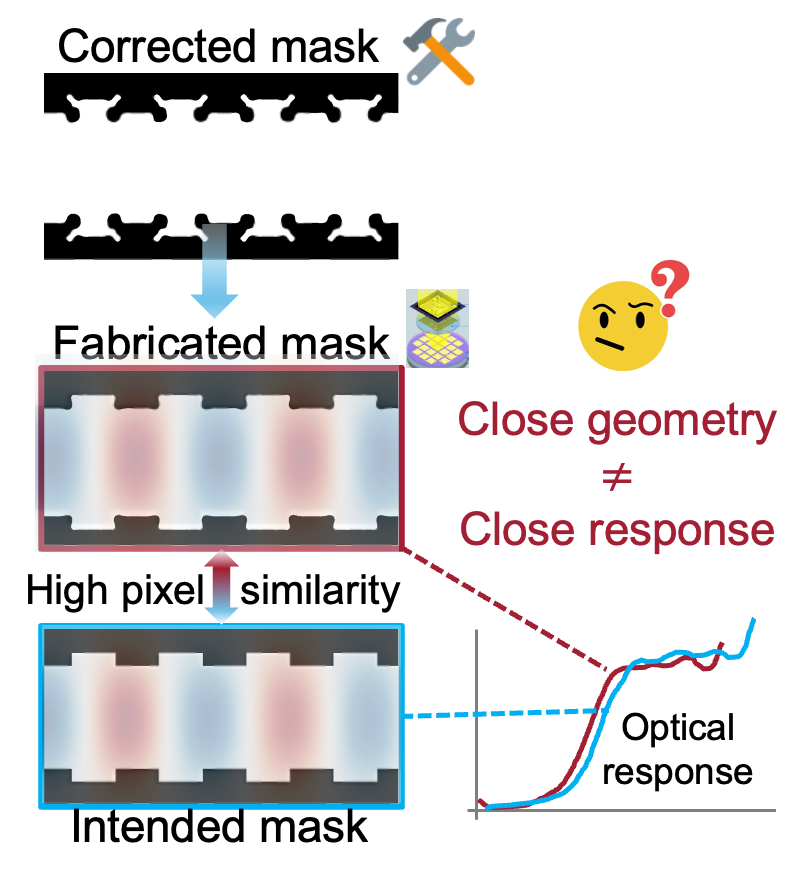}
    \label{fig:Motivation_Sensitive}
    }
    \hspace{8pt}
    \subfloat[]{\includegraphics[width=0.32\linewidth]{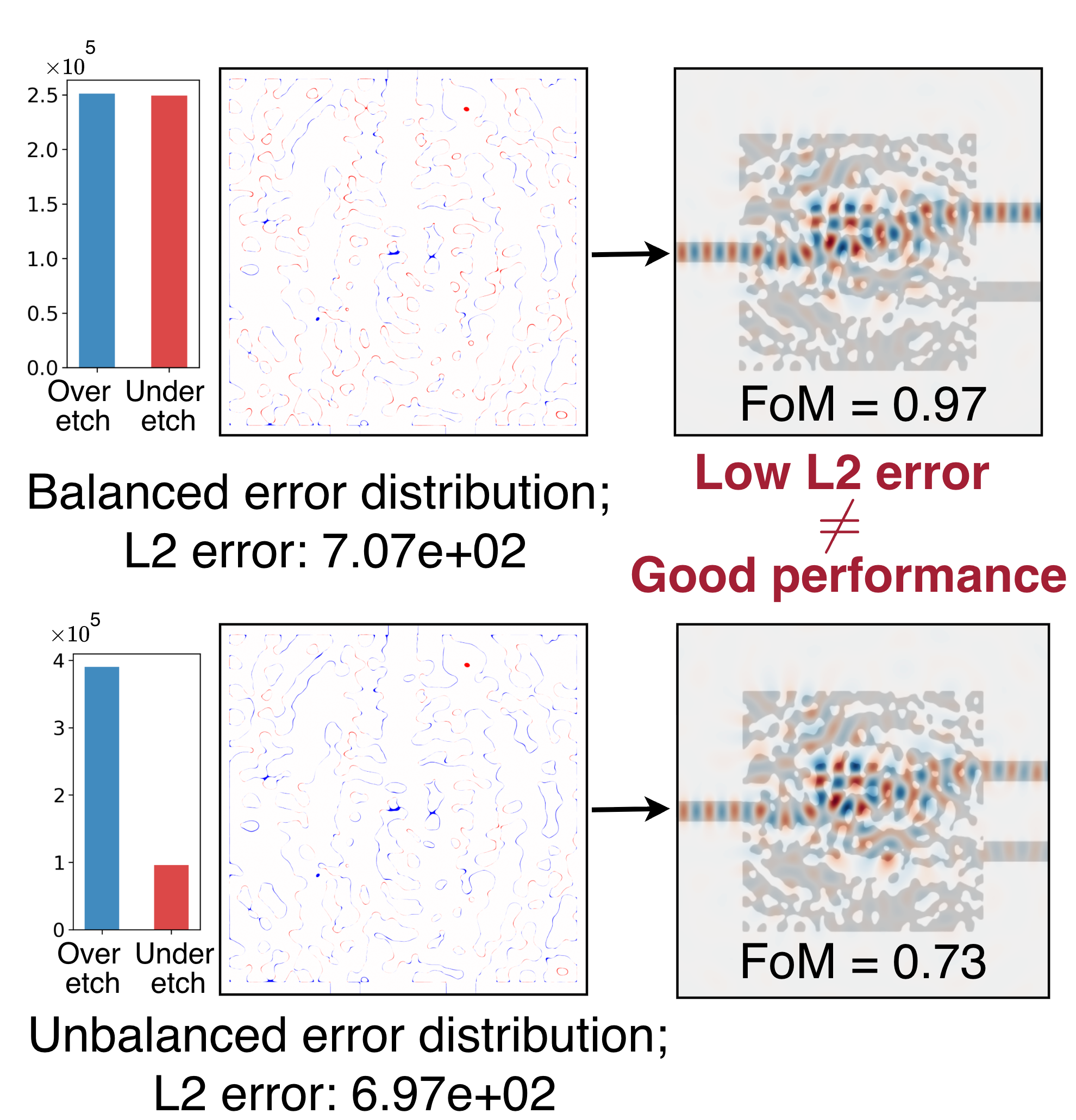}
    \label{fig:Motivation_L2}
    }
     \hspace{5pt}
    \subfloat[]{\includegraphics[width=0.31\linewidth]{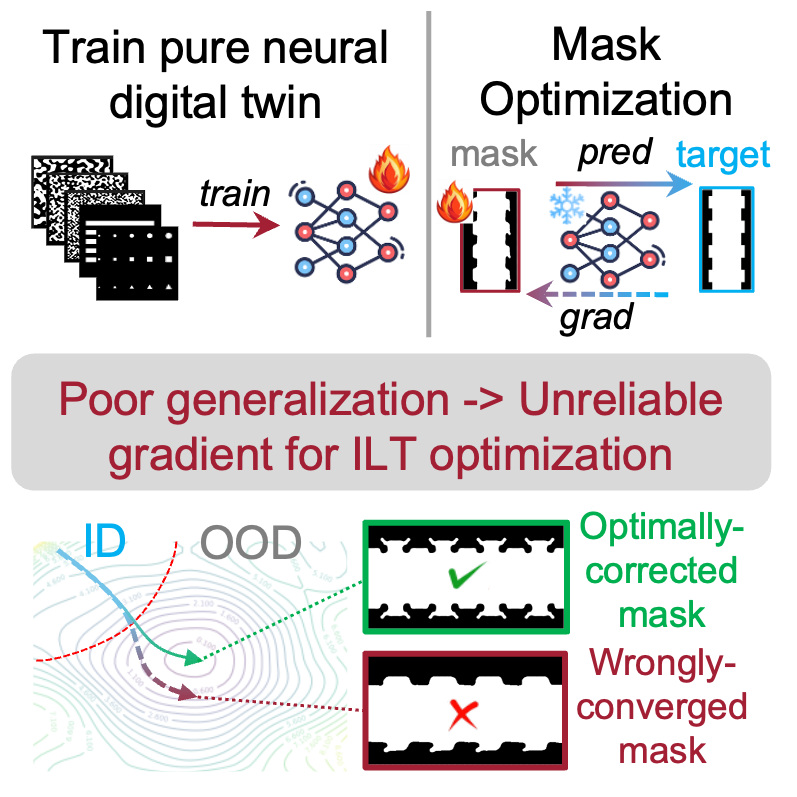}
    \label{fig:Motivation_GradErr}
    }
    \vspace{-10pt}
    \caption{Comprehensive motivation. 
    (a) Standard mask correction is insufficient, as \emph{close geometry} does not always lead to close optical response.
    (b) Low pixel-wise $\ell_2$ error does not necessarily imply good photonic performance. An \emph{unbalanced} error distribution dominated by systematic over-/under-etch introduces a global bias in geometry and can degrade the FoM significantly, despite a comparable $L_2$ value. 
    (c) Neural network fabrication model trained on calibration data might have poor generalization to unseen patterns and mislead ILT optimization.}
    \label{fig:Motivation2}
    \vspace{-10pt}
\end{figure*}

%% file: doc/2_prelim.tex
\section{Preliminary}
\label{sec:Preliminaries}

\subsection{Photonic Inverse Design and Fabrication Awareness}
Photonic inverse design has emerged as a powerful paradigm for synthesizing compact, high-performance photonic components by directly optimizing electromagnetic objectives, enabling structures that outperform conventional hand-designed devices.
A growing body of work has extended inverse design to account for fabrication constraints, including minimum feature size control~\cite{NP_chen2020design, NP_gershnabel2022reparameterization, hammond2022high, NP_khoram2020controlling}, smoothing or regularization of design variables, and robustness-aware objectives that mitigate sensitivity to process variation. These approaches have improved the practicality of inverse-designed devices by shaping the design space itself.
Our work is \emph{complementary rather than competing with fabrication-aware inverse design (FAID)}~\cite{piggott2017fabrication}. We assume a given inverse-designed layout, produced by any existing design methodology, and focus on the downstream manufacturing mask optimization.
\name can be combined with FAID to further improve yield, or applied independently to existing inverse-designed components without redesign.
To motivate mask optimization, we briefly summarize the manufacturing steps that shape the printed pattern and how they are typically modeled. Figure~\ref{fig:Fabrication_process} sketches a standard pattern-transfer flow.

\subsection{Fabrication Process Modeling}
\input{figtex/fig_fabrication_process}

In modern chip fabrication, a complete process flow often consists of hundreds of steps. Patterned masks are transferred to the wafer through a sequence of processes, including lithography, development, ion implantation, etching, and polishing, all of which collectively affect the final pattern geometry on the wafer. Among these steps, lithography is one of the most critical stages because it largely determines the achievable resolution and is therefore crucial to both chip performance and manufacturing yield. In this section, we briefly introduce several key processes and discuss how they are commonly modeled and simulated in practice. Throughout this paper, we use lowercase letters for scalars, bold lowercase for vectors, $\odot$ for element-wise production, $\otimes$ for convolution, $^{*}$ for complex conjugate, $M$, $I$, $Z$, $W$ for mask, aerial, resist and final wafer image, $\mathcal{F}$ and $\mathcal{F}^{-1}$ for fast Fourier transform (FFT) and inverse FFT, respectively.

\noindent\textbf{Photolithography Process}.~ The lithography simulation models are designed to mimic the printing effects without performing actual lithography. In practice, the Hopkins diffraction model~\cite{banerjee2013iccad} of the partial coherence imaging system is widely adopted to approximate the printing behavior of the lithography. Its purpose is a separation of the influence of the mask and the imaging system, including the pupil function and the illumination. For numerical convenience, the Hopkins method is often stated in terms of the spatial spectrum of the aerial intensity $I$:
\begin{equation}
\small
\mathcal{F}(I)(f,g) =
\iint_{-\infty}^{\infty}
\mathcal{T}\bigl((f'+f,\; g'+g),\; (f',\; g')\bigr)\,
\mathcal{F}(M)(f'+f,\; g'+g)\,
\mathcal{F}(M)^{*}(f',\; g')\,
df'\, dg'
\label{lab:equ_I}
\end{equation}
\begin{equation}
\small
\mathcal{T}\bigl((f',\,g'),\,(f'',\,g'')\bigr)
:= 
\iint_{-\infty}^{\infty}
\mathcal{F}(J)(f,g)\,
\mathcal{F}(H)\bigl(f+f',\, g+g'\bigr)\,
\mathcal{F}(H)^{*}\bigl(f+f'',\, g+g''\bigr)\,
df\, dg
\label{lab:equ_tcc}
\end{equation}
where $\mathcal{T}$ is the Transmission Cross-Coefficient (TCC), the weighting factor $J$ depends only on the effective source, and $H$ denotes the projector transfer function. From Eqs.~\eqref{lab:equ_I} and ~\eqref{lab:equ_tcc}, it can be seen that $\mathcal{T}$ is independent of the mask transmission function $\mathcal{F}(M)$.

To reduce the computational overhead, a singular value decomposition (SVD) approximation is typically adopted for the Hopkins model, called the sum of coherent sources (SOCS)~\cite{Cobb1995SumOC}. The basic idea is to take the SVD of the coefficient matrix. The TCC spectrum matrix can be written as $\mathcal{F}(\mathcal{T}) = \sum_{i} \alpha_i\, h_i\, h_i^{*}$, where $h_i$ are the eigenvectors and $\alpha_i$ are the eigenvalues of the TCC spectrum $\mathcal{F}(\mathcal{T})$. Applying the inverse Fourier transform then yields aerial intensity by the SOCS formulation:
\begin{equation}
\label{eq:SOCS}
I = \sum_{i} \alpha_i \left|\mathcal{F}^{-1}\!\left(\mathcal{F}(h_i)\odot \mathcal{F}(M)\right)\right|^{2}.
\end{equation}

\noindent\textbf{E-beam Lithography Process}.~ 
Unlike photolithography, electron-beam lithography (EBL) is a maskless direct-write process whose pattern fidelity is mainly governed by electron--matter scattering in the resist/substrate stack (including forward- and back-scattering), which leads to the well-known proximity effect~\cite{zhou2006monte}. 
Therefore, EBL process modeling typically starts from a physical scattering model, where Monte Carlo simulation is widely used to track electron trajectories in the resist and substrate and to estimate the spatial distribution of deposited energy. 
For computational efficiency, the Monte Carlo results are often summarized by a point spread function (PSF), which is fitted using parametric models such as a double-Gaussian~\cite{gentili1990energy} to capture both short-range forward scattering and long-range backscattering contributions:
\begin{equation}
\mathrm{PSF}(r)
=
\frac{\eta}{\pi \alpha^{2}}
\exp\!\left(-\frac{r^{2}}{\alpha^{2}}\right)
+
\frac{1-\eta}{\pi \beta^{2}}
\exp\!\left(-\frac{r^{2}}{\beta^{2}}\right),
\label{eq:psf_double_gaussian}
\end{equation}
where $r$ denotes the radial distance from the beam incidence point, $\alpha$ and $\beta$ are the characteristic scattering ranges (Gaussian widths) for the forward- and back-scattering components, respectively, and $\eta\in[0,1]$ is the weight of the forward-scattering term (thus $1-\eta$ corresponds to the backscattering weight). The normalization factors ensure that each Gaussian term integrates to its corresponding weight over the 2D plane, so that the total PSF integrates to unity.
Given an incident dose map derived from the layout, the energy deposition density can then be computed by convolving the dose map with the PSF.

\noindent\textbf{Development Process}.~After exposure, development models predict the final resist profile (e.g., critical dimension, sidewall angle, resist loss) from the aerial image. A widely used approach is to convert the aerial image into a latent image (effective exposure in resists)~\cite{dill1975characterization}, then use a rate-based dissolution model~\cite{thackeray2007chemically} to describe how fast each resist location dissolves in developer. In practice, this is often implemented as: (1) a fixed or variable threshold/contour model~\cite{randall1999variable} for fast approximation; or (2) a more accurate dissolution-rate model where the local dissolution speed depends on exposure level, capturing effects such as resist contrast, diffusion/blur, and post-exposure bake smoothing.

\noindent\textbf{Etching Process}.~Given the developed resist pattern, etch modeling describes how it transfers into films, predicting etch depth, CD bias, and profile distortions. 
In most process simulators, etching is formulated as a moving-boundary/surface-evolution problem~\cite{economou2000modeling}, where the material surface advances at a local etch rate determined by simplified plasma physics and feature geometry.

\noindent\textbf{Deposition Process}.~After pattern transfer, deposition models~\cite{lundback2011modelling} predict conformality and topography growth on 3D features (e.g., step coverage and gap fill). Most models combine a global deposition rate with geometry-dependent effects (orientation, shadowing, and transport/reaction limits in narrow trenches), and iteratively update the surface to estimate post-deposition thickness nonuniformity.

\noindent\textbf{Polishing Process}.~To recover planarity for subsequent layers, chemical–mechanical polishing (CMP) models~\cite{lee2016mechanical} predict post-polish topography (e.g., dishing and erosion), which is critical for multilayer patterning and lithography focus control. Typical models combine a global removal component (pressure, speed, slurry) with layout-dependent effects driven by local pattern density; simulators often compute an effective local contact/pressure from a windowed density map and iteratively update the surface to obtain post-CMP thickness and planarity metrics.

\subsection{Inverse Lithography in EDA and Uniqueness in Photonic ILT}
The lithography process is a critical step in chip manufacturing, as it transfers the mask pattern onto the wafer and largely determines the critical dimension, performance, and yield. 
In EDA, DFM techniques such as optical proximity correction and ILT are mature and widely deployed, forming a closed-loop interface between design and manufacturing.
Recent work has leveraged data-driven and neural surrogates~\cite{chen2020damo, ye2019lithogan, liu2023adversarial, lin2018data} to accelerate lithography modeling and enable gradient-based ILT~\cite{yang2018gan, jiang2020neural, zhu2023l2o, chen2023develset, yang2022generic, chen2024differentiable} under increasingly complex process conditions~\cite{sun2024efficient}. These methods have been highly effective for rectilinear, logic-style layouts and geometry-centric objectives.

However, photonic ILT introduces several unique challenges and motivations.
(i) Inverse-designed photonic layouts are often \emph{not manufacturing-feasible} under production DUV nodes: severe diffraction-driven distortions (rounding, blurring, bridging) can cause dramatic FoM degradation and a heavy worst-case tail, leading to near-zero yield for some devices.
(ii) Unlike logic OPC, where geometric fidelity is a strong proxy, photonic performance is governed by nonlocal electromagnetic interactions; thus, \textbf{``close geometry'' does not imply ``close function.''} Small but structured errors or systematic biases (e.g., global over-/under-etch) can dominate optical response even when pixel-wise error is low.
(iii) Accurate lithography modeling remains expensive and highly process-specific. 
Detailed foundry information (e.g., calibrated optical process parameters and OPC recipes) is often proprietary and not readily accessible to designers, \emph{especially when most photonic foundries do not provide OPC/ILT services}. 
This creates a major barrier to designing high-performance, compact, inverse-designed photonic devices, which are typically highly sensitive to fabrication imperfections. 
The lack of a reliable process model also hinders fabrication-aware inverse design, making model acquisition a central challenge. 
A faithful model benefits both sides: it helps foundries diagnose and improve processes, enhance yield, and develop advanced PDKs, while enabling designers to produce robust photonic devices under realistic manufacturing non-ideality.

To address the aforementioned problem,
\name brings the power of ILT to integrated photonic circuits with photonics-specific adaptation by incorporating physics/optics in the data synthesis, model design, and mask optimization workflow, bridging the simulation-fabrication gap in photonic inverse design through a photonics-aware design-technology co-optimization loop.

%% file: figtex/fig_fabrication_process.tex
\begin{figure}
    \centering
    \includegraphics[width=0.9\columnwidth]{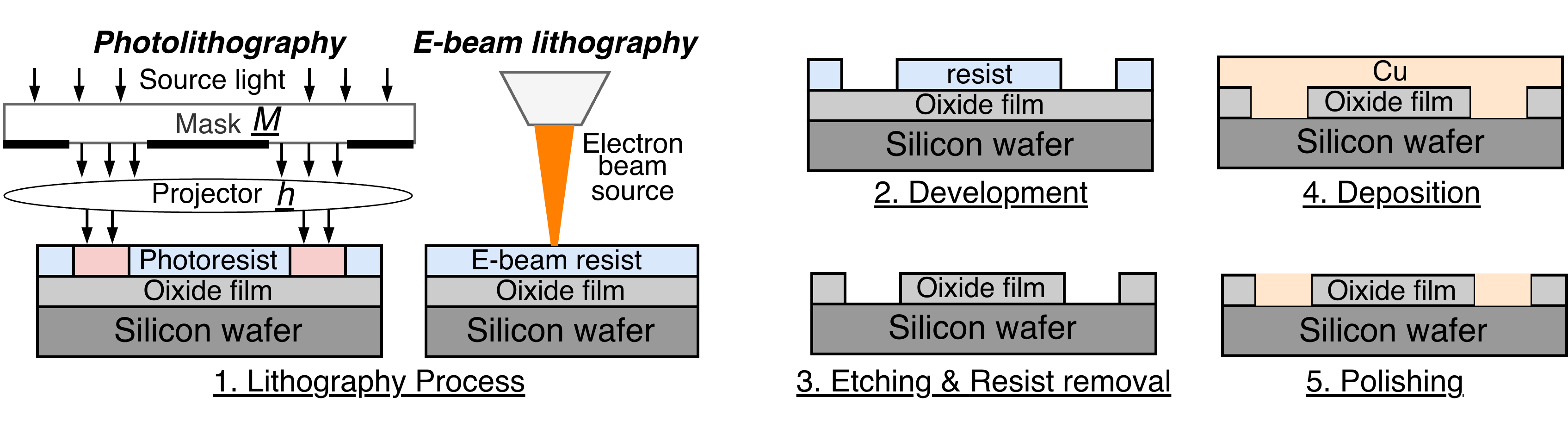}
    \vspace{-5pt}
    \caption{Illustration of a typical pattern transfer flow in chip fabrication. Patterns are first defined by either photolithography or E-beam lithography to form a resist image on an oxide-coated silicon wafer, followed by (2) resist development, (3) etching and resist removal to transfer the pattern into the underlying film, (4) material deposition (e.g., Cu fill), and (5) chemical-mechanical polishing to planarize the surface.
    }
    \label{fig:Fabrication_process}    
\end{figure}

%% file: doc/3_method.tex
\section{Proposed \name Framework}
\label{sec:Method}

\input{figtex/fig_flow}

Inverse-designed photonics breaks two assumptions that standard EDA ILT relies on: (i) mask geometric fidelity is the sole objective, and (ii) process modeling can be treated as a solved, foundry-provided service. 
In photonics, neither holds. 
Optical function is governed by a small set of sensitivity-critical structures, so minimizing pixel-wise geometry error can still yield a large performance drop; meanwhile, many photonic designers do not have access to calibrated OPC recipes or high-volume process characterization. 
\name is designed around a single goal: \textbf{turn photonic curvilinear mask optimization into a data-efficient, stable, and optical function-preserving design-technology co-optimization loop that both photonic designers and foundries can benefit from low-cost yield augmentation}.

Figure~\ref{fig:framework} encapsulates the key paradigm of \name to achieve this goal: close the loop between a locally calibratable fabrication digital twin and a physics-guided inverse mask optimization engine, so that photonic ILT can become a first-class, accessible new design stage introduced into general yield-oriented PIC design closure.

The remainder of this section details the three pillars of Fig.~\ref{fig:framework}.
In Sec.~\ref{sec:calibration_data}, we synthesize a calibration set for identifiability under budget constraints: a compact pattern dataset that prioritizes curvilinear motifs and proximity interactions that are most informative for inverse-designed photonics, maximizing process learnability per unit cost in fabrication, metrology, and labeling.
In Sec.~\ref{sec:fabrication_model}, we learn a physically-grounded, parametric fabrication digital twin that captures complicated stochastic fabrication processes and, crucially, provides stable gradients to guide inverse mask optimization.
Finally, Sec.~\ref{sec:ilt_method} closes the loop by integrating the differentiable fabrication digital twin into a physics-guided photonic ILT procedure that operates in latent topology space, enforcing manufacturability while explicitly protecting performance-critical features under process variations, beyond deterministic pixel-wise geometry matching.
The ILT-corrected layouts are then used for downstream simulation, verification, and yield assessment, and can be iteratively refined, together with the fabrication digital twin model, enabling accelerated design convergence.

\vspace{-5pt}
\subsection{Curvilinear Calibration Pattern Synthesis for Photonic ILT}
\label{sec:calibration_data}
\input{figtex/fig_model_training}
A practical bottleneck for photonic ILT is \textbf{calibration}: \emph{learning an accurate fabrication digital twin from limited post-fab measurements}, since chip fabrication area, metrology (e.g., SEM imaging), and label annotation are all expensive.
\name formulates calibration as an identifiability-driven pattern synthesis problem, we aim to select a compact set of layouts whose fabricated outcomes maximally reflect the dominant fabrication distortions encountered in inverse-designed photonic masks.

\noindent\textbf{Complementary Pattern Synthesis Strategies in \name}.~
\name provides three calibration pattern categories, each targeting a distinct aspect of the process response:

\ding{202}~\uline{Regular patterns as controlled probes}: We include both rectilinear and curvilinear photonic motifs (lines/slots, bends, tapers, junctions, corners) and their compositions. These patterns provide \emph{interpretable} responses and isolate specific mechanisms (e.g., linewidth bias, corner rounding, curvature-dependent bias), making them efficient anchors for model learning.

\ding{203}~\uline{Randomized freeform patterns as spectral coverage probes}: Inverse-designed photonics contains rich multi-scale structure; capturing this requires coverage across spatial frequencies and feature sizes. We therefore generate randomized curvilinear patterns with controlled power spectra to span a broad frequency distribution, from smooth low-frequency contours to high-frequency edge/corner detail, while varying feature widths and spacings. 
These patterns efficiently expose nonlocal proximity interactions and process nonlinearities that are difficult to excite using uncontrolled random patterns.

\ding{204}~\uline{Inverse-designed device crops as distribution-matching probes}: Finally, we include patterns sampled directly from inverse-designed devices (or cropped regions of interest). This aligns calibration with the true data distribution seen at deployment, reducing domain shift and improving on-task accuracy when the goal is yield augmentation for inverse-designed devices.

\noindent\textbf{Pattern Layout Arrangement}.~
All patterns are tiled into a compact calibration reticle with controlled spacing.
As shown in Fig.~\ref{fig:Calibration_Data}, each pattern occupies a $3~\mu\mathrm{m}\times 3~\mu\mathrm{m}$ window, and adjacent patterns are separated by a $2~\mu\mathrm{m}$ spacing to mitigate mutual interference (e.g., proximity and loading effects). 
The pattern size is chosen based on the SEM field-of-view (typically a few microns), so that each SEM image can fully capture one pattern reticle, thereby reducing the need for precise alignment during SEM image stitching.

\noindent\textbf{"Virtual Foundry" as Controlled Study Vehicle}.~
To systematically evaluate \name across pattern choices and model designs, we create a \emph{virtual foundry} as a controlled proxy to represent an arbitrary fabrication flow. 
The virtual foundry serves as a synthetic fabrication data generator/oracle, producing paired mask–wafer examples with adjustable process settings and variations, allowing us to (i) study generalization under limited calibration, (ii) isolate how calibration distributions affect downstream ILT, and (iii) conduct algorithmic ablation studies that would be prohibitively expensive in real fabrication.
Importantly, the virtual foundry is not a claim about any specific foundry; it is a \emph{blackbox hidden testbed} for stress-testing the proposed calibration methodology and quantifying design-process co-optimization behavior.

\uline{For the DUV fabrication setting}, we emulate the photolithography step using the sum of coherence system (SOCS) representation of the Hopkins imaging model, a standard abstraction in computational lithography that is both physically grounded and computationally efficient. 
Specifically, we construct the DUV lithography model using \textit{Torchlitho}~\cite{chen2024open}, an open-source framework for computing the transmission cross-coefficient (TCC) kernel from the illumination source $J$ and the projector system transfer function $H$. The model parameters, including the wavelength, numerical aperture (NA), and related process parameters, are calibrated based on preliminary SEM data collected from our previous AIM Photonics MPW SiPho chip tape-out fabricated with 193-nm DUV lithography. The calibrated Hopkins-based model fits the annotated SEM labels reasonably well, with acceptable residual errors, suggesting that it captures the dominant lithography-induced distortions observed in real foundry data to some extent.
Using the decomposed SOCS kernels, we then evaluate the aerial image intensity via Eq.~\eqref{eq:SOCS}. 
We obtain the developed resist pattern by thresholding the aerial image, followed by a parametrized pattern-transfer surrogate that emulates common non-ideal fabrication effects, including shot-to-shot dose variation, stochastic resist/etch defects, resolution-limited line bridging, and curvature-dependent etching threshold variation.
Concretely, we apply a set of morphological and geometric operators that form a minimal “distortion basis.”
We remove small isolated resist islands, discarding foreground components whose area is below a preset limit. It models stochastic resist residues and defect-like speckles that would not reliably transfer through etching. 
We then apply the morphological \emph{closing} operation to seal narrow gaps and form unintended connections, mimicking \emph{bridging} phenomena caused by insufficient resolution, resist scumming, or etch microloading. 
Next, we apply the morphological \emph{opening} operation to eliminate tiny protrusions and thin connections, which approximates the breakage of fragile features and the removal of narrow resist necks during pattern transfer. 
Finally, we model systematic curvature-dependent bias (e.g., corner rounding and over-/under-etch) by computing a signed distance field and offsetting the effective etch front by a curvature-dependent shift, so highly curved boundaries experience stronger local bias than near-straight segments.

\uline{For the EBL fabrication setting}, the exposure step is modeled by an energy deposition map produced by electron forward scattering in the resist and long-range backscattering from the substrate. 
The Monte-Carlo scattering results can be summarized by a parametric point spread function (PSF). 
Specifically, we use the \emph{three-Gaussian plus exponential} (\textit{3G+exp}) model~\cite{liu2021hnu}:
\begin{equation}
\small
\label{eq:ebl_psf_3gexp}
P(r)=\frac{1}{\pi(1+\eta+\eta'+\eta'')}
\left[
\frac{1}{\alpha^2}\exp\!\left(-\frac{r^2}{\alpha^2}\right)
+\frac{\eta}{\beta^2}\exp\!\left(-\frac{r^2}{\beta^2}\right)
+\frac{\eta'}{\gamma^2}\exp\!\left(-\frac{r^2}{\gamma^2}\right)
+\frac{\eta''}{2\gamma_2^2}\exp\!\left(-\frac{r}{\gamma_2}\right)
\right],
\end{equation}
where $r$ denotes the distance from an exposure point; $\alpha$ is the forward-scattering range; $\beta$ is the backscattering range; $\gamma$ captures mid-range scattering; and $\gamma_2$ controls the exponential tail. The dimensionless ratios $\eta$, $\eta'$, and $\eta''$ weight the backscattering, mid-range, and exponential components relative to the forward-scattering term, respectively.
Given a (pixelized) dose map $d(\mathbf{r})$, the deposited energy map $E(\mathbf{r})$ is computed by convolving $d$ with the PSF:
\begin{equation}
E(\mathbf{r}) = (P\otimes d)(\mathbf{r}).
\label{eq:ebl_energy_conv}
\end{equation}
We calibrated these parameters by fitting the
outputs of the PreFab model~\cite{prefab}, which was trained on real
EBL fabrication data. Specifically, we queried PreFab with a diverse set of input patterns
and used the resulting input-output pairs to tune our EBL virtual foundry.
We then reuse the same virtual-fabrication pipeline as in the DUV setting (development + pattern-transfer surrogate), with the resist threshold set to 0.5 to obtain the developed resist pattern.

Across both DUV and EBL, these parameterized operators provide a controllable source of nonlocal blur and nonlinear stochastic pattern-transfer effects, enabling thorough studies and yield assessment without requiring repeated physical fabrication. 
It should be noted that the current virtual foundry is intended as a feasibility investigation focusing on the dominant and practically important sources of fabrication distortion, particularly lithography and etching/development effects. As more SEM images become available and more comprehensive models of additional fabrication processes, such as development, etching, polishing, and layout-dependent process effects, are incorporated, the virtual foundry can be further refined to improve its predictive accuracy.

\subsection{Learning a Differentiable Fabrication Digital Twin}
\label{sec:fabrication_model}
The photonic inverse lithography procedure in \name flow embeds the fabrication digital twin inside a gradient-based inverse mask optimization loop.
Therefore, the digital twin must satisfy a stronger requirement than forward-only predictors: beyond accurate wafer image prediction on nominal (uncorrected) masks, it must provide \textbf{physically consistent and stable gradients with respect to mask perturbations, and remain reliable on the out-of-distribution, ILT-corrected masks encountered during optimization}.

\noindent\textbf{Digital Twin Architecture Design}.~
Because lithography largely determines pattern fidelity and introduces the dominant non-locality in printing, we adopt a \emph{factorized} differentiable fabrication model consisting of (i) a \emph{lithography module} that maps the mask $M$ to a continuous exposure field (i.e., aerial image in DUV lithography or the energy deposition map in EBL), and (ii) a \emph{thresholding module} that summarizes downstream development/etch transfer by projecting this field into a binary (or soft-binary) wafer pattern.
This interpretable decomposition injects physically meaningful prior knowledge, reduces learning difficulty under limited calibration data, and yields reliable gradients for the downstream ILT loop.

Formally, given an intended design mask $M\in\mathbb{R}^{S_i\times S_i}$, the predicted wafer image $W\in\mathbb{R}^{S_o\times S_o}$ is obtained as follows,
\begin{equation}
W=\Phi_{\theta = (\theta_I,\theta_T)}(M)
=\psi\!\bigl(I_{\theta_I}(M)-T_{\theta_T}(M)\bigr),
\ \ 
\psi(x)=\frac{\tanh(2\beta x)+1}{2},
\label{eq:fab_model_output}
\end{equation}
where $I_{\theta_I}(M)\in\mathbb{R}^{S_o\times S_o}$ denotes the predicted continuous exposure field produced by the lithography module with learnable parameters $\theta_I$, and $T_{\theta_T}(M)\in\mathbb{R}^{S_o\times S_o}$ denotes a threshold map produced by the threshold module with learnable parameters $\theta_T$, and $\psi(\cdot)$ is an element-wise differentiable projection function with binarization sharpness controlled by $\beta$.

Motivated by the need for both accurate forward prediction and reliable gradients in ILT, we study two complementary parameterizations of the lithography module $I_{\theta_I}$: 
(1) a \textbf{data-driven neural twin} that uses a U-Net backbone to learn the multi-scale nonlocal mask-to-exposure mapping; and
(2) a \textbf{physics-parameterized twin} that instantiates $I_{\theta_I}$ with a physics-grounded DUV/EBL model with learnable optical parameters, e.g., Zernike coefficients in DUV SOCS kernels or variables in a double-Gaussian EBL PSF function, to capture system imperfections.

We \textbf{hypothesize} that, while a data-driven neural twin may achieve competitive forward prediction accuracy with enough calibration data, a physics-grounded twin will offer better generalization and reliable gradient in the data-limited regime, particularly for challenging DUV lithography and for the out-of-distribution (OOD) masks produced during iterative ILT.
In the later evaluation section, we evaluate both digital twin designs, comparing not only wafer prediction accuracy but also downstream ILT correction quality and yield.

\noindent\textbf{Physics-Grounded Data-Efficient Learning}.~
To fully leverage the collected and annotated fabrication data, we partition each labeled mask–wafer pair into many training samples via \emph{patch-based supervision}: we extract densely overlapping patches with a sliding window and apply fabrication-invariant augmentations (translations, rotations, and flips), which preserve lithography physics while improving data efficiency.

Lithography processes are intrinsically nonlocal; each pixel is coupled to its neighborhood.
Naïvely training on isolated patches introduces artificial boundary conditions: pixels near the crop boundary lack the surrounding context required to predict the correct exposure/printing outcome, leading to biased learning and unreliable artifacts.
To respect this physics, each training sample uses an \emph{expanded input} of size $S_i=\texttt{patch\_size}+\texttt{context\_size}$, while supervision is applied only on the center cropped region in the output of size $S_o=\texttt{patch\_size}$, as illustrated in Fig.~\ref{fig:Model_Training}.
Equivalently, the model predicts on the full field, but we treat only the valid interior region as trustworthy and discard boundary pixels. The \texttt{context\_size} is chosen to exceed the effective receptive field of the model, ensuring that every supervised output pixel is computed from sufficient neighborhood information.

We train the fabrication digital twin by minimizing the mean square error (MSE) loss between the prediction $W$ and the labeled wafer image $W^{*}\in\mathbb{R}^{S_o\times S_o}$:
\begin{equation}
\mathcal{L}_{\mathrm{MSE}}(W,\,W^{*})
=\frac{1}{N\,S_o^{2}}
\sum_{n=1}^{N}
\left\|\Phi_{\theta}\!\left(M^{(n)}\right)-W^{*(n)}\right\|_2^2,
\label{eq:general_loss}
\end{equation}
where $N$ is the number of training samples. 

For computational efficiency in both training and downstream ILT, we optionally downsample the patch resolution from the native $1000~\mathrm{px}/\mu\mathrm{m}$ in the GDS mask, e.g., to $250~\mathrm{px}/\mu\mathrm{m}$, trading spatial detail for speed with minimal impact on predicted pattern fidelity in our study.

\subsection{Photonics-Informed Inverse Lithography for Mask Correction}
\label{sec:ilt_method}
\input{tables/alg_ilt}
Photonic ILT \emph{differs from traditional pattern-matching in one key respect}: the objective is not merely to reproduce the nominal geometry pixel-by-pixel, but to recover the subset of structures that most strongly govern optical function under variations, while maintaining mask printability under a given process.
Algorithm~\ref{alg:ilt} describes this differentiable, photonics-informed optimization loop by combining (i) a well-conditioned levelset-based topology parameterization, (ii) explicit manufacturability priors, and (iii) performance-aware weighting derived from adjoint sensitivity.

\input{figtex/fig_ilt_flow}
\noindent\textbf{Manufacturability-aware Levelset Topology Parameterization}.~
Directly optimizing the full-resolution binary mask is poorly conditioned: it encourages high-frequency pixel artifacts and yields unmanufacturable masks.
Given the target design mask $M_0\in[0,1]^{H_0\times W_0}$ at the original resolution, we initialize a latent levelset field $\rho_0\in\mathbb{R}^{H_0\times W_0}$ around zero according to the nominal inverse designed mask $M_0$, so that foreground and background are represented by small positive (0.1) and negative values (-0.1), respectively. 
To improve optimization efficiency, instead of optimizing $\rho_0$ at the original resolution, we downsample the latent levelset knots to a lower resolution and optimize
$\rho\in\mathbb{R}^{H\times W}$ as the latent topology variable:
\begin{equation}
\rho = \mathcal{D}\!\left(\rho_0; s\right)
= \mathrm{Interp}_{\mathrm{bilinear}}\!\left(\rho_0;\, s\right),
\ \  s=r_{\ell}/{r_0},
\label{eq:p_downsample}
\end{equation}
where $\mathcal{D}(\cdot)$ denotes the bilinear interpolation operation, $r_0$ and $r_{\ell}$ are the original and latent resolutions (in $\mathrm{px}/\mu\mathrm{m}$), and $s$ is the corresponding scale factor.

After obtaining the latent levelset knots $\rho$, at each optimization iteration, we first apply a Gaussian blur to enforce spatial smoothness of the levelset function and suppress high-frequency artifacts (e.g., pixel-level jaggies and unrealistically sharp corners), which also encourages manufacturable feature sizes.
We then interpolate this levelset function back to the original mask resolution using bilinear interpolation and convert it to a differentiable mask via a sigmoid-based soft binarization:
\begin{equation}
\begin{aligned}
\hat{M} = \mathcal{D}\!\left(p \otimes G;\, s\right), \quad s={r_0}/r_{\ell},\quad M = 1/\big(1+\exp(-\alpha\,\hat{M})\big),
\end{aligned}
\label{eq:mask_reparam}
\end{equation}
where $G$ is a Gaussian smoothing kernel, $\alpha$ is the sharpness factor (set to $200$ in this work). 
By construction, $M(x,y)\in[0,1]$. 
Since our fabrication prediction model takes inputs of size $S_i\times S_i$, we further slice $M$ into partially overlapping patches before prediction, as shown in Fig.~\ref{fig:ILT_flow}. 
Specifically, we extract $B$ patches and group them into mini-batches for parallel inference, denoted by $M_B\in\mathbb{R}^{B\times S_i\times S_i}$.
The model then produces the corresponding batch of wafer patches
:
\begin{equation}
    W_B=\frac{1}{3}\sum_{v_i\in v}\Phi_{\theta}(M_B;\, \beta, v_i)
    \label{eq:ilt_prediction}
\end{equation}
where $\beta$ is the inference-time sharpness in Eq.~\eqref{eq:fab_model_output} (fixed to $s=20$ in our experiments), and $v$ is the process-variation vector (nominal + corner cases ($v_{min}, v_{nom}, v_{max}$)=(0.85, 1.0, 1.15)), and we scale the $T_{\theta_T}$ output, thereby providing a simple way to emulate dose/resist/etch variations in the fabrication process.
We then stitch all predicted patches back to the full predicted wafer image $W\in[0,1]^{H_0\times W_0}$ by averaging in the overlapping regions and calculate the ILT loss. 

\noindent\textbf{Photonics-Informed ILT Objective Beyond Pixel-wise Matching}.~
Finally, we optimize the latent variable by minimizing an $\ell_4$ loss guided by a user-specified photonic-informed weight map $\Gamma\in\mathbb{R}^{H_0\times W_0}$:
\begin{equation}
\mathcal{L}_{\mathrm{4}}(W, W^{*})
=||\Gamma \odot (W-W^{*})||_{4}^4.
\label{eq:l4_loss}
\end{equation}
In this study, we guide ILT using the normalized \emph{adjoint sensitivity} map of the device figure of merit, i.e., $\Gamma=|\partial\text{FoM}/\partial M_0|$ (normalized to [0.5, 1.5]).
Intuitively, the adjoint sensitivity identifies mask regions where small geometric perturbations induce the largest FoM change, and thus where faithful recovery matters most.
This \textbf{embeds photonic insight (design-intent) directly into ILT}: the optimizer concentrates correction effort on performance-critical features, while allowing FoM-insensitive regions to relax strict pixel-wise fidelity and instead deform to accommodate neighboring corrections, improving overall FoM-driven parametric yield. 
Additionally, the adjoint sensitivity is precomputed only once on the intended target design mask and applied as a weighting map to the loss calculation. The additional computational cost mainly comes from an element-wise weighting operation, whose overhead is negligible.

Finally, the corrected mask is legalized through morphology operations (opening and closing) to help satisfy the mask-rule check (MRC) and improve mask printability. 
However, this post-processing step does not guarantee the removal of all rule violations, especially when the inverse-designed device itself contains intrinsically small or closely spaced features that violate the target PDK design-rule check (DRC). 
Incorporating DRC/MRC fixing directly into the inverse-design or ILT loop is an important future research topic and is beyond the main scope of this work.

%% file: figtex/fig_flow.tex
\begin{figure}
    \centering
    \includegraphics[width=\columnwidth]{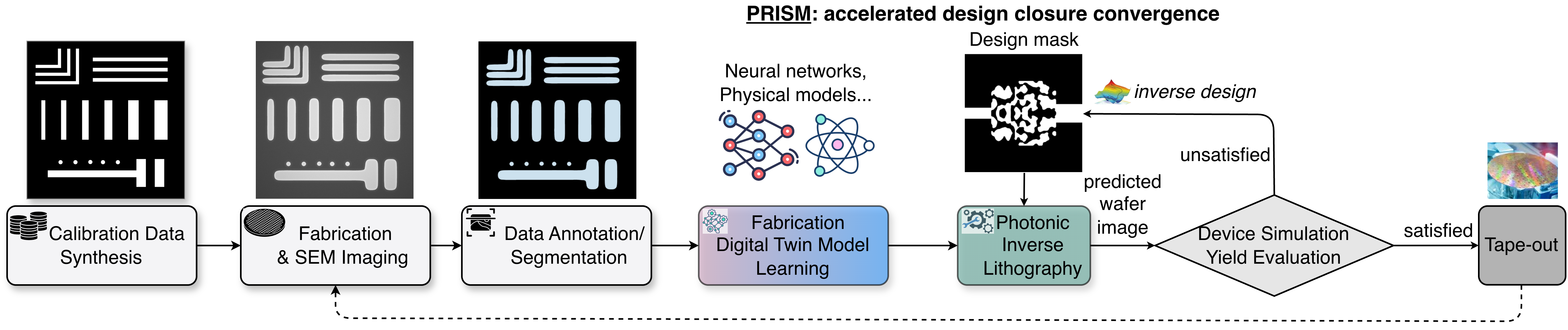}
    \vspace{-10pt}
    \caption{Overview of proposed \name flow to accelerate photonic inverse design convergence.
    }
    \vspace{-8pt}
    \label{fig:framework}    
\end{figure}

%% file: figtex/fig_model_training.tex
\begin{figure*}
    \centering
    \subfloat[]{\includegraphics[width=0.45\linewidth]{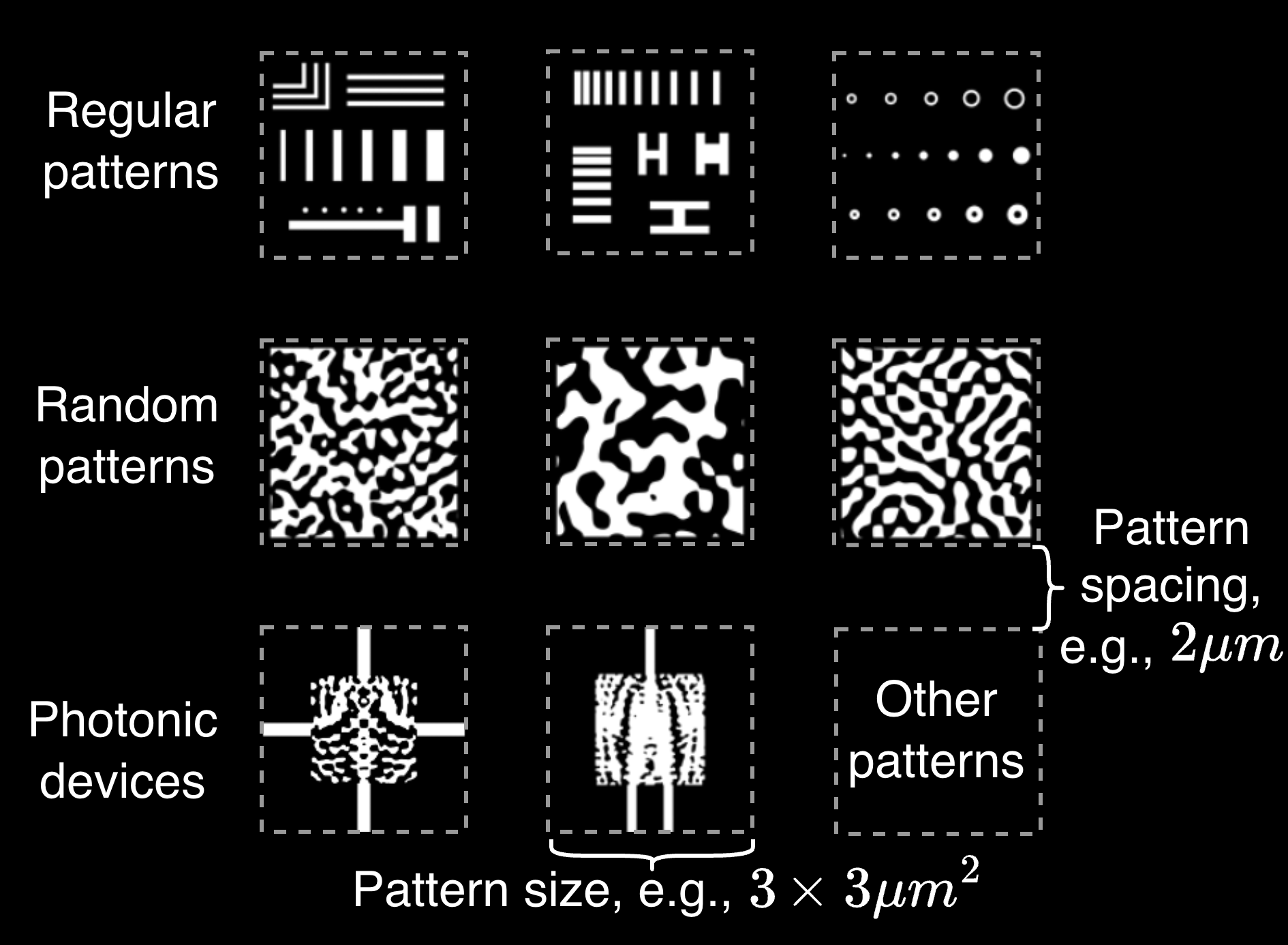}
    \label{fig:Calibration_Data}
    }
    \hspace{8pt}
    \subfloat[]{\includegraphics[width=0.44\linewidth]{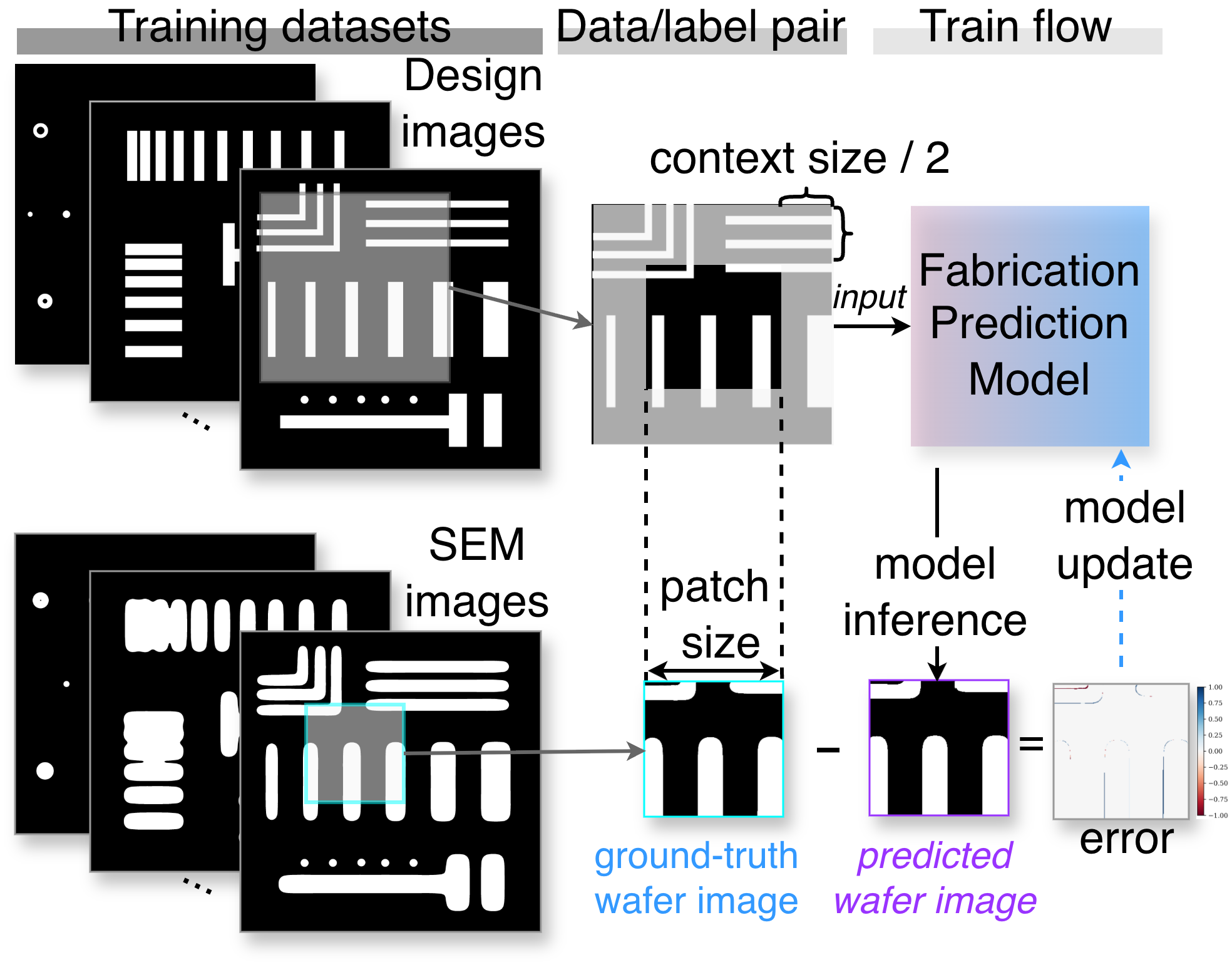}
    \label{fig:Model_Training}
    }
    \vspace{-5pt}
    \caption{(a) Illustration of regular, random, and photonic device patterns for fabrication model learning.
    (b) Design images and post-fab wafer images form data/label pairs for fabrication model training. }
    \vspace{-10pt}
    \label{fig:MotivationDataset_Train}
\end{figure*}

%% file: tables/alg_ilt.tex
\begin{algorithm}[t]
\caption{Photonics-informed, variation-aware ILT in latent topology space}
\label{alg:ilt}
\begin{algorithmic}[1]
\State \textbf{Input:} target mask $M_0\in[0,1]^{H_0\times W_0}$, target wafer label $W^{*}$ (or desired wafer target),
fabrication twin $\Phi_\theta(\cdot;\beta,v)$, weight map $\Gamma\in[0,1]^{H_0\times W_0}$
\State \textbf{Hyperparameters:} latent resolution $r_\ell$, original resolution $r_0$, sharpness $\alpha$, blur kernel $G_{\mathrm{blur}}$,
inference sharpness $\beta$, process variation scale $v=(v_{min}, v_{nom}, v_{max})$, step size $\eta$, iterations $K$
\State \textbf{Output:} corrected mask $M$

\vspace{0.3em}
\State \textbf{Initialize latent field at full resolution.}
\State $\rho_0 \in \mathbb{R}^{H_0\times W_0} \gets 0.1\cdot \mathbbm{1}[M_0>0.5] - 0.1\cdot \mathbbm{1}[M_0\le 0.5]$
\State Downsample to latent topology variable:
\State $\rho \gets \mathcal{D}(p_0; s),\ \ s = r_\ell/r_0$ \Comment{Eq.~\eqref{eq:p_downsample}}

\vspace{0.3em}
\For{$k = 0$ to $K-1$}
    \State \textbf{(1) Mask printability constraints.}
    \State $\tilde{\rho} \gets \rho \otimes G_{\mathrm{blur}}$ \Comment{Gaussian blur in latent space}

    \State \textbf{(2) Project latent variable to a differentiable mask.}
    \State $\hat{M} \gets \mathcal{D}(\tilde{\rho}; s),\ \ s = r_0/r_\ell$ \Comment{upsample, Eq.~\eqref{eq:mask_reparam}}
    \State $M \gets \bigl(1+\exp(-\alpha\,\hat{M})\bigr)^{-1}$ \Comment{sigmoid re-parameterization, Eq.~\eqref{eq:mask_reparam}}

    \State \textbf{(3) Patch-based forward prediction (context-aware).}
    \State Extract $B$ overlapping input patches $M_B \in \mathbb{R}^{B\times S_i\times S_i}$ from $M$ \Comment{Fig.~\ref{fig:ILT_flow}}
    \State Predict wafer patches with variation injection $v$: $W_B \gets \frac{1}{3}\sum_{v_i\in v}\Phi_\theta(M_B;\beta,v_i)$ \Comment{Eq.~\eqref{eq:ilt_prediction}}
    \State Stitch $W_B$ into full wafer $W\in[0,1]^{H_0\times W_0}$ by averaging overlaps

    \State \textbf{(4) ILT objective and update.}
    \State $\mathcal{L} \gets \left\| \Gamma \odot (W - W^{*}) \right\|_{4}^4$ \Comment{weighted $\ell_4$, Eq.~\eqref{eq:l4_loss}}
    \State $\rho \gets \rho - \eta \,\nabla_{\rho}\mathcal{L}$ \Comment{backprop through stitching and $\Phi_\theta$ and apply update}
\EndFor
\State \textbf{return} $M$
\end{algorithmic}
\end{algorithm}

%% file: figtex/fig_ilt_flow.tex
\begin{figure}
    \centering
    \includegraphics[width=0.9\columnwidth]{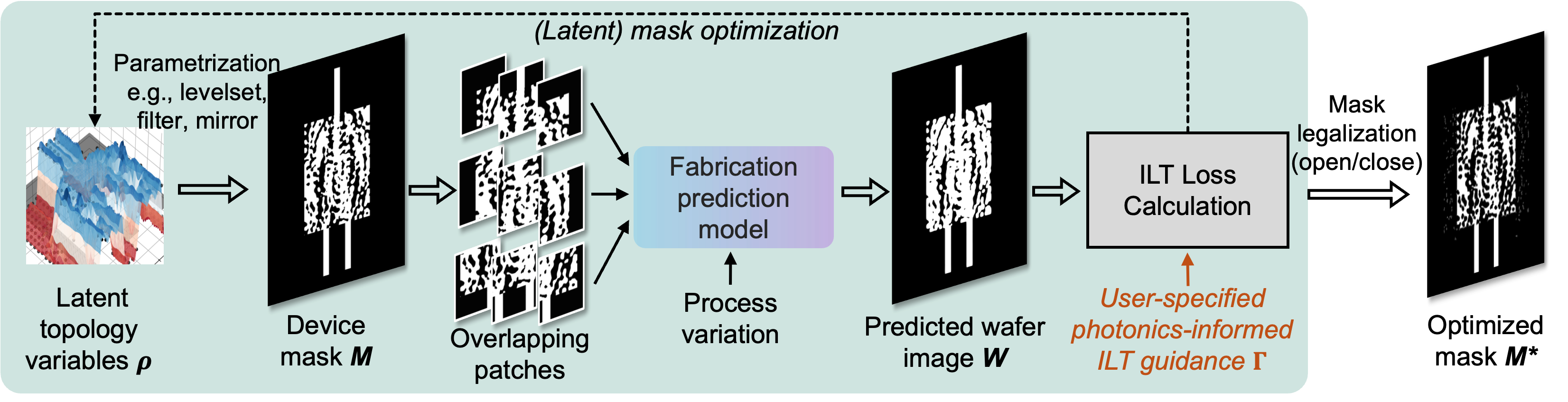}
    \vspace{-5pt}
    \caption{Our proposed photonic inverse lithography flow performs variation-aware, photonics-informed mask optimization.
    }
    \vspace{-5pt}
    \label{fig:ILT_flow}    
\end{figure}

%% file: doc/4_result.tex
\section{Evaluation Results}
\label{sec:Results}
This section evaluates \name along two coupled axes: \textbf{(i) fabrication digital twin prediction accuracy} and \textbf{(ii) downstream effectiveness for photonic ILT and yield recovery}. 
While wafer-image prediction accuracy is a necessary prerequisite, ILT places a stricter requirement on the gradient reliability of the digital twin. 
We therefore evaluate both forward prediction quality and robustness on ILT-generated masks, and then provide insights on how different model designs and device properties impact end-to-end ILT effectiveness.
Our experiments are conducted on a Linux server with a 64-core AMD EPYC 9554 CPU (base frequency $\sim$ 3.1\, GHz) and four NVIDIA RTX PRO 6000 Blackwell GPUs, each with 98GB VRAM.

\noindent\textbf{Digital Twin Training Dataset}.~
In practice, SEM imaging and reliable labeling are the dominant bottlenecks for calibration. 
Hence, we focus on learning high-performing digital twin predictors under a constrained data volume budget. 
Following Sec.~\ref{sec:calibration_data}, we synthesize three calibration datasets that reflect complementary coverage of the mask space: \emph{regular patterns} (interpretable regular motifs), \emph{frequency-diverse random patterns}, and \emph{inverse-designed device samples}. 
These choices allow us to study the cost-accuracy trade-off of calibration, and to identify which pattern families best support robust digital twin learning.
Specifically, for both the regular pattern and random pattern datasets, we select $10$ patterns of size $3~\mu\mathrm{m}\times3~\mu\mathrm{m}$ as the training set. 
We place patterns with a $2~\mu\mathrm{m}$ spacing to mitigate pattern-to-pattern proximity interference, resulting in an \emph{effective patterned area} of $10\times(3\times3)=90~\mu\mathrm{m}^2$, while the \emph{total occupied area} of the calibration block is approximately $(5\times5)\times10=250~\mu\mathrm{m}^2$. 
Similarly, we select representative photonic-device samples with a comparable effective patterned area ($\approx 90~\mu\mathrm{m}^2$) to match the calibration budget, including a bend, a Bragg grating, a $1\times3$ splitter, an MDM, and a TDM.

\input{figtex/fig_device}
\noindent\textbf{Digital Twin Test Dataset}.~
To assess both wafer image prediction accuracy and generalization on ILT masks, our test set comprises two subsets: (\textbf{Test 1}) a photonic-device dataset containing inverse-designed devices and several regular, process-variation-sensitive photonic structures, and (\textbf{Test 2}) an ILT-corrected photonic device dataset. 
Notably, the inverse-designed devices in the test set are generated using different random seeds from those used in the photonic-device training set, ensuring non-overlapping design instances. 
In addition, Test~1 further includes extra device families beyond the training distribution, such as the 2.5D silicon devices shown in Fig.~\ref{fig:device}.
For Test 2, we directly apply established physics-based ILT pipelines to generate corrected masks: the ICCAD-2013 SOCS kernels~\cite{banerjee2013iccad} for DUV and a PSF-based model for EBL~\cite{gentili1990energy}. 
Note that these pipelines are different from (and not calibrated against) our virtual foundry settings, as \emph{our virtual fab is strictly hidden}.
Evaluating these ILT-corrected masks allows us to probe model robustness on mask distribution drift induced by ILT.

\noindent\textbf{Evaluation Metrics for Wafer Image Prediction}.~
Given the predicted wafer image $W$ and the ground-truth wafer image $W^{*}$, let $N$ denote the total number of pixels. We define MSE, pixel accuracy (PA), and intersection-over-union (IoU) as
\begin{equation}
    \label{eq:Metrics}
    \mathrm{MSE} = \frac{1}{N}\left\|W^{*}-W\right\|^{2};\quad
    \mathrm{PA}  = \frac{W^{*}\cap W}{W^{*}};\quad 
    \mathrm{IoU} = \frac{W^{*}\cap W}{W^{*}\cup W},
\end{equation}
where $\cap$ and $\cup$ denote pixel-wise intersection and union.

\noindent\textbf{Fabrication Digital Twin Model}.~
\uline{\textbf{CNN}}.~
We adopt a CNN~\cite{purwono2023understanding} architecture similar to~\cite{gostimirovic2022deep}, but remove the fully connected layers and keep only convolutional layers. It is a lightweight fully-convolutional model with 5 blocks of (3$\times$3 Conv + BN + ReLU) and channel widths $\{32,64,128,128,128\}$ followed by a $1\times1$ convolution and a sigmoid output, totaling $\approx0.4$M parameters.
\uline{\textbf{UNet}~\cite{ronneberger2015u}}.~
A lightweight UNet is implemented (depth=3, base channel=8) with bilinear upsampling and skip connections, using double 3$\times$3 conv blocks per stage and a \textbf{3$\times$3 + 1$\times$1 + Sigmoid} head, totaling $\approx0.2$M parameters.
\uline{\textbf{FNO}~\cite{li2020fourier}}.~ The model combines a reduced Fourier neural operator branch (average pooling by 16, up to 32$\times$32 Fourier modes, 64-channel output) with a shallow spatial convolutional branch (two 3$\times$3 convolutions), then upsamples and concatenates the features and predicts the output with a 3$\times$3 + 1$\times$1 convolution plus sigmoid, for five convolution layers plus one Fourier layer and roughly $\approx0.4$M parameters.
\uline{\textbf{Learnable-SOCS for DUV}}~\cite{Cobb1995SumOC}.~
We assume the imaging conditions are known (typical DUV $\lambda=193\,\mathrm{nm}$ and $\mathrm{NA}=1.35$) as well as the illumination source. The pupil function is modeled with a fixed amplitude defined by the NA cutoff and a \emph{learnable} phase parameterized by a small set of Zernike coefficients. 
The only learnable parameters are the Zernike coefficient vector $\theta=\mathbf{c}\in\mathbb{R}^{N_Z}$ (with $N_Z$ basis functions, $N_Z=12$), which defines the pupil phase aberration and is optimized from data, while the source map and NA-limited amplitude remain fixed. 
Given the source support, we construct the SOCS/TCC operator by stacking source-weighted, shifted pupil functions into a matrix and performing a direct SVD decomposition to obtain a low-rank set of SOCS kernels (top singular vectors) and their weights (squared singular values). 
These (data-calibrated) SOCS kernels are then applied in the Fourier domain to efficiently simulate aerial images for ILT.
\uline{\textbf{Double-Gaussian PSF model for EBL}~\cite{gentili1990energy}}.~
For EBL, we use a physics-motivated double-Gaussian PSF model and approximate the aerial pattern by convolving the mask with a learnable PSF. The learnable parameters include the mixture weight parameter $\eta$ (mapped to a valid nonnegative mixing ratio) and four inverse standard deviations $\sigma_{ax}^{-1}, \sigma_{ay}^{-1}, \sigma_{bx}^{-1}, \sigma_{by}^{-1}$. Physically, the first Gaussian captures the narrow/core blur of the beam, while the second Gaussian models a broader tail due to long-range scattering/proximity effects; $\eta$ controls their relative contribution, and the $\sigma$ parameters determine the blur strength and anisotropy along the two axes. 

\subsection{Evaluation of Fabrication Digital Twins on Wafer Image Prediction}

\input{tables/tab_ebl_corrected}

\input{tables/tab_duv_corrected}
Table~\ref{tab:ebl_modeling} and Table~\ref{tab:duv_modeling} compare wafer-image prediction accuracy for the EBL and 193nm DUV processes. We evaluate both (i) uncorrected device masks (\textbf{Test1}) and (ii) ILT-corrected masks produced during optimization (\textbf{Test2}), since the latter reflects the out-of-distribution inputs that a digital twin must handle when embedded in the ILT loop.

\noindent\textbf{EBL vs. DUV}.~ 
Across all methods, \textbf{EBL is substantially easier to model} than 193nm DUV: on both Test1 and Test2, all candidates achieve near-saturated segmentation metrics (PA/IoU typically $\approx 0.998$-$0.999$) with low regression error (MSE $\sim 10^{-3}$). 
In contrast, \textbf{DUV prediction is more challenging}: for CNN and FNO, MSE increases by roughly an order of magnitude relative to EBL (from $\sim10^{-3}$ to $\sim10^{-2}$), and IoU drops noticeably (down to $\approx 0.87$-$0.96$). 
This gap is consistent with the stronger diffraction-limited resolution and longer-range interactions in DUV, which induce more nonlinear and nonlocal behavior.

\noindent\textbf{Model Robustness and OOD Generalization}.~
Within each process, \emph{UNet and the physics-grounded digital twins are the most robust}, while CNN and FNO are less stable, especially on DUV and on ILT-corrected inputs. 
Concretely, on DUV Test1, UNet/SOCS achieve MSE $\sim(3$-$4)\times10^{-3}$ with IoU $\approx 0.985$-$0.988$, whereas CNN/FNO exhibit markedly larger errors. 
On the more challenging DUV Test2, CNN/FNO degrade further (IoU $\approx 0.87$--$0.91$, MSE up to $\sim4\times10^{-2}$), whereas UNet remains comparatively stable. 
Overall, these trends suggest that (i) strong multi-scale spatial modeling (UNet) improves prediction stability, and (ii) embedding physical structure (PSF/SOCS) provides additional robustness and generalizability when the input mask distribution shifts due to ILT.

\noindent\textbf{Impacts of Calibration Dataset on Model Generalization.}~
Finally, we evaluate how calibration pattern choice impacts generalization to various device structures and ILT-corrected masks. 
Overall, all three datasets synthesized within the \name framework (regular, frequency-diverse random patterns, and device samples) provide sufficiently informative coverage to learn the core mask-to-wafer mapping and achieve strong performance on both Test1 and Test2. 
Differences across datasets are modest and scenario-dependent: frequency-diverse random patterns can improve robustness to distribution shift in some cases, while regular primitives offer a particularly favorable cost-performance trade-off due to their ease of fabrication, metrology, and annotation. 
Collectively, these results validate the \name calibration strategy: a compact, budget-aware pattern set can support accurate digital twin learning and generalization to the mask distributions encountered in downstream ILT.

\subsection{Evaluation of \name on Photonic Inverse Lithography}
\label{sec:ilt_result}

\input{tables/tab_ebl_ilt}
\input{tables/tab_duv_ilt}
\input{figtex/fig_wdm_plot}

\input{figtex/fig_grating}

In the ILT experiments, we evaluate post-fabrication photonic performance under both EBL and DUV processes, comparing \emph{uncorrected designs} against \emph{ILT-corrected masks}, where each mask is optimized for 100 iterations. 
For each device, we measure the transmission at each output port, e.g., $|S_{21}|^2$ for the transmission from port~1 to port~2, and characterize the distribution of post-fabrication performance under process variations.

\noindent\textbf{Process Variation Sampling in Virtual Foundry.}~
We model fabrication variations by sampling the virtual-fab parameters from Gaussian distributions and generating an ensemble of post-fabrication wafer images.
Let $\xi$ denote a random variation instance, $W(\xi)$ the corresponding fabricated wafer pattern, and $\mathcal{F}(\cdot)$ a photonic figure of merit such as $|S_{ij}|^2$.
We draw $\xi\sim\mathcal{N}(0,\Sigma)$ and evaluate $\mathcal{F}(W(\xi))$ over multiple samples.
Specifically, the sampled parameters are
\begin{equation}
\small
\begin{aligned}
\textbf{Dose:}\quad 
& d(\xi)\sim \mathcal{N}\!\left(d_0,\ (0.02\,d_0)^2\right), \\[2pt]
\textbf{EBL lithography:}\quad 
& \alpha(\xi)\sim \mathcal{N}\!\left(\alpha_0,\ (0.05\,\alpha_0)^2\right), \\[2pt]
\textbf{DUV lithography:}\quad 
& h(\xi)\sim \mathcal{N}\!\left(0,\sigma_h^2\right),\qquad \sigma_h=25, \\[-1pt]
& K_{\mathrm{SOCS}}(\xi)=K_{\mathrm{SOCS}}\!\bigl(h(\xi)\bigr), \\[2pt]
\textbf{Post-lithography:}\quad
& t(\xi)\sim \mathcal{N}\!\left(t_0,\ (0.02\,t_0)^2\right), \\
& k_{\mathrm{open}}(\xi)\sim \mathcal{P}_{\mathrm{open}},\quad 
  k_{\mathrm{close}}(\xi)\sim \mathcal{P}_{\mathrm{close}},\quad
  \tau_{\kappa}(\xi)\sim \mathcal{P}_{\kappa}, \\[2pt]
\textbf{Ensemble:}\quad
& \{W(\xi_i)\}_{i=1}^{N_s},\qquad 
  \{\mathcal{F}(W(\xi_i))\}_{i=1}^{N_s}.
\end{aligned}
\label{eq:virtualfab_sampling}
\end{equation}
where $d_0$ is the nominal exposure dose, $\alpha_0$ is the nominal forward-scattering parameter in the EBL PSF model (Eq.~\eqref{eq:ebl_psf_3gexp}), $h$ denotes the defocus value used to generate SOCS kernels $K_{\mathrm{SOCS}}(\xi)$ in DUV, and $t_0$ is the nominal resist threshold in the threshold stage. 
Moreover, $k_{\mathrm{open}}$ and $k_{\mathrm{close}}$ are the structuring-element sizes used in morphological open/close operations, and $\tau_{\kappa}$ denotes the curvature-based etching threshold; $\mathcal{P}_{\mathrm{open}}$, $\mathcal{P}_{\mathrm{close}}$, and $\mathcal{P}_{\kappa}$ represent user-specified sampling distributions (e.g., uniform sampling). 

For each sample $\xi_i$, we (i) draw dose $d(\xi_i)$ and process-specific lithography parameters, (ii) generate the aerial image/energy deposition map, and (iii) draw post-lithography parameters (e.g., $t(\xi_i)$, $k_{\mathrm{open}}(\xi_i)$, $k_{\mathrm{close}}(\xi_i)$, $\tau_{\kappa}(\xi_i)$) to obtain the post-fabrication wafer image $W(\xi_i)$ via the virtual-fab pipeline. 
We then evaluate the device performance $\mathcal{F}(W(\xi_i))$ over $\{W(\xi_i)\}_{i=1}^{N_s}$ to estimate the post-fabrication performance distribution and yield.

\noindent\textbf{Yield Definition.}~
Given a target performance value $f^*$ (e.g., the original design's $|S_{ij}|^2$) and sampled post-fab performance
$f_i=\mathcal{F}(W(\xi_i))$, we define the \emph{parametric yield} at level $p\in(0,1)$ as the fraction of samples achieving at least a $\rho$-relative performance:
\begin{equation}
\mathrm{yield}_{p}
=\frac{1}{N_s}\sum_{i=1}^{N_s}\mathbf{1}\!\left[f_i \ge p\, f^*\right].
\end{equation}
For example, $\mathrm{yield}_{80\%}$ counts the percentage of samples whose post-fabrication performance is no worse than $80\%$ of the target.

\noindent\textbf{\name-based ILT Result Summary.}~
Table~\ref{tab:ebl_ilt} and Table~\ref{tab:duv_ilt} report ILT results on both inverse-designed photonic devices and regular yet process-sensitive devices under EBL and 193nm-DUV processes. 
Based on the digital-twin study in Sec.~\ref{sec:fabrication_model}, we report results using the best-performing twins for each setting: \textbf{\name-based ILT (UNet)} and \textbf{\name-based ILT (PSF)} for EBL, and \textbf{\name-based ILT (SOCS)} for DUV.

Across devices, \textbf{EBL} exhibits mild process-induced degradation for the evaluated structures, and both \name-based ILT (UNet) and \name-based ILT (PSF) often achieve comparable yield improvements. 
This suggests that, for these device classes, the EBL fabrication process is easier to learn, making both data-driven and physics-parametrized digital twins effective engines in ILT.

In contrast, under \textbf{DUV} the uncorrected designs can suffer severe post-fabrication degradation, particularly for highly resonant grating-like geometries (e.g., WDM mux / Bragg grating), where $\mathrm{yield}_{80\%}$ / $\mathrm{yield}_{90\%}$ can collapse to near zero in many cases. 
In this regime, \name-based ILT substantially improves yield, with \textbf{\name-based ILT (SOCS)} consistently providing the most reliable recovery in the most process-sensitive cases.
Critically, the DUV results \uline{support our hypothesis} that, under limited calibration data, a learnable physics-grounded digital twin generalizes more reliably and provides more stable gradients for the ILT loop, particularly on the out-of-distribution masks generated during iterative correction, whereas \emph{over-parameterized neural twins can achieve good forward prediction yet fail to provide a reliable first-order oracle for effective ILT optimization}. Figure~\ref{fig:ilt_mask} visualizes \name-based ILT (SOCS) outcomes under 193\, nm DUV, showing the original design masks $M_0$, the optimized masks $M$, and the corresponding change maps $M-M_0$, which highlight the structured, physically meaningful mask adjustments introduced during correction.

\noindent\textbf{Case Study 1: Inverse-designed WDM mux}.~
Figure~\ref{fig:wdm_plot} contrasts the WDM device under \textbf{EBL} and \textbf{DUV} in the nominal setting, showing the input mask, virtual-fabricated wafer image, and resulting device performance.
Under \textbf{EBL}, ILT requires only modest boundary refinements.
Both UNet and PSF variants achieve strong recovery.
Under \textbf{DUV}, however, the correction demands substantial mask modifications to counter diffraction-limited imaging and long-range proximity effects, and \name-based ILT (SOCS) provides more reliable recovery than UNet thanks to its superior generalization and reliable gradient flow due to explicit encoding of diffraction physics.

\noindent\textbf{Case Study 2: Spectrum-sensitive Bragg grating}.~
Figure~\ref{fig:grating_plot} highlights a complementary regime: the Bragg grating is structurally regular, yet its spectral response is highly sensitive to small fabrication perturbations due to its long dimension that accumulates error and resonant working principle.
Under 193 nm DUV, diffraction-induced blur and long-range proximity effects distort the effective index and duty cycle, directly perturbing the coupling coefficient and accumulated phase, which produces large passband/stopband shifts.
The mask/wafer/spectrum triplet in Fig.~\ref{fig:grating_plot} illustrates why a physics-grounded digital twin is critical for gradient-based ILT: \name-based ILT (UNet) exhibits large post-fabrication deviation from the intended topology, suggesting optimization drift into an OOD mask regime, whereas \name-based ILT (SOCS) produces more structured corrections with interpretable assist-like features and yields higher-fidelity post-fab wafer patterns with improved duty-cycle and teeth height recovery.
Consistent with this, SOCS-based ILT yields spectra that more closely track the target response under sampled process variations.

\noindent\textbf{Case Study 3: Variation-aware ILT and Adjoint Gradient-based ILT Guidance}.~
\input{figtex/fig_yield_optimization}
Figure~\ref{fig:AdjointYield} presents an ablation study validating two mechanisms for improving robustness: \textbf{(i) variation-aware ILT}, where the ILT objective is optimized over three representative process corners (min/nominal/max), and \textbf{(ii) adjoint-guided ILT}, where a normalized adjoint sensitivity map reweights the geometry loss to emphasize performance-critical regions. 
Compared to nominal-only ILT without weighting, each mechanism improves the post-fabrication performance distribution, and their combination yields the most consistent gains, improving average device FoM while reducing tail risk under process variations.

\noindent\textbf{Case Study 4: ILT-Challenging Instances Under Variation (WDM and TTS)}.~
Despite these improvements, a small subset of devices remains challenging to fully recover the yield under significant DUV process variation.
In particular, the WDM and thermally tunable switch (TTS) still exhibit limited yield even with SOCS-based ILT, e.g., $\mathrm{yield}_{80\%}=54\%$ for the WDM mux and $\mathrm{yield}_{80\%}=24\%$ for TTS. 
These are the hardest cases in our benchmark because they exhibit intrinsic sensitivity due to strong scattering and resonance, where fabrication errors accumulate over long optical paths and induce nontrivial drift on their sharp spectra.
These cases point to an \uline{important insight}: fully closing the design-fabrication gap for inverse designed photonics may require moving beyond post hoc ILT on fixed designs toward \emph{fabrication-aware inverse design (FAID) from the outset}. 
ILT primarily seeks to recover the original design intent under a given process model, but it cannot fundamentally reduce a device’s intrinsic sensitivity to process variations.
Fabrication-aware inverse design (FAID) is an orthogonal direction that improves robustness \emph{by construction}. 
In the following section, we further show that FAID with explicit minimum-feature-size control can work synergistically with \name, producing masks that are easier to correct and yielding improved post-fabrication performance after ILT.

\input{figtex/fig_ilt_mask}

\vspace{-5pt}
\subsection{"FAID-ILT Synergy": Impacts of Device Area and MFS on Yield and ILT Effectiveness}
Building on the ILT study in Sec.~\ref{sec:ilt_result}, we now shift perspective from ILT algorithm-centric evaluation to a device-centric question: \emph{which device geometries are intrinsically more recoverable under fabrication distortions, and how does this interact with ILT?} While ILT can compensate for systematic process bias, its effectiveness depends strongly on the geometric characteristics of the underlying design.
To answer this, we systematically study a family of $1\times2$ WDM multiplexers with varying \emph{design region area} and \emph{minimum feature size (MFS)} constraints. 
The design region area trades off device compactness and available functional degrees of freedom, while MFS (enforced via Gaussian-smoothing during levelset-based inverse design) limits the smallest printable features, representing a fabrication-aware inverse-design (FAID) approach.

\input{tables/tab_device_area}
Table~\ref{tab:mfs} reveals a three-way trade-off between \textbf{design freedom (area)}, \textbf{nominal performance (FoM)}, and \textbf{fabrication robustness / ILT recoverability}:
\uline{Smaller design regions} reduce available degrees of freedom and thus lower the best attainable nominal FoM under ideal fabrication.
\uline{Increasing MFS} further restricts the effective design space, typically degrading the peak nominal FoM by suppressing fine-scale features.
However, for \uline{sufficiently large design regions} (e.g., $8\times8~\mu\mathrm{m}^2$), \emph{increasing MFS substantially improves post-fabrication yield after DUV ILT, even when nominal FoM decreases slightly}.
In particular, larger MFS values (e.g., $\mathrm{mfs}=300$) applied in FAID eliminate fragile sub-resolution features. 
As a result, the generated masks lie closer to a ``printable space'' of the DUV process, making the ILT correction target more achievable.
These results suggest a practical guideline for fabrication-aware photonic inverse design under DUV: \uline{Allocate sufficient design area to preserve functional degrees of freedom, and impose a moderate-to-large minimum feature size so that the geometry stays in a manufacturable regime where ILT remains effective}.
In this view, FAID and ILT are complementary: FAID improves robustness \emph{by construction}, while ILT compensates residual systematic distortions. 
Together, they yield designs that are both high-performing and reliably recoverable under process variations.

\subsection{\name-based ILT on Complicated-Function SiN Inverse-Designed Photonic Devices}
\input{tables/tab_sin}
\input{figtex/fig_sin}

To validate the scalability of \name{} and to examine the process sensitivity of low-index-contrast silicon nitride (SiN) devices, we perform mask ILT on two representative SiN components: a $1\times5$ MDM mux ($17~\mu\mathrm{m}\times9~\mu\mathrm{m}$) and a $1\times4$ WDM mux ($24~\mu\mathrm{m}\times24~\mu\mathrm{m}$). 

Table~\ref{tab:sin_ilt} summarizes the nominal-case results under both EBL and 193\, nm DUV virtual fabrication. Under the EBL process, the post-fabrication device performance remains close to the ideal design for both devices, reflecting the high pattern fidelity of EBL. Under the DUV process, the MDM mux still exhibits only a moderate degradation, whereas the WDM mux suffers a severe performance drop after fabrication. This discrepancy is mainly attributed to the grating-based (pseudo-periodic) filtering mechanism of the WDM, whose spectral response is highly sensitive to duty-cycle errors, linewidth bias, and edge-placement distortions, as visualized in Fig.~\ref{fig:SiN}. Encouragingly, applying \name-based ILT (SOCS) substantially reduces the printed-pattern error and largely restores the desired WDM transmission response.
These results suggest that low-index-contrast devices with broader modal profiles (e.g., MDMs) tend to be intrinsically more robust to DUV process distortions, while grating-like, phase-sensitive structures (e.g., WDMs) require stronger lithography-aware correction; in such cases, a physics-grounded predictor enables more reliable recovery under DUV.

\subsection{\name-based ILT on 3D Inverse-Designed Devices}
To further demonstrate \name’s applicability beyond 2.5D benchmarks, we evaluate an inverse-designed \emph{3D Si vertical grating coupler} that couples an out-of-plane Gaussian beam into the in-plane fundamental mode, effectively acting as a \emph{photonic via} for future 3D very large-scale photonic integration (VLPI).
This device is very challenging to design in such a compact footprint, making inverse design an essential enabler technique.
Figure~\ref{fig:gc3d} reports the nominal-case 193nm DUV virtual fabrication results using 3D FDTD simulation, comparing the uncorrected mask and \name-based ILT (SOCS).
In the ideal simulation, the coupler achieves a peak coupling efficiency of 0.394.
Under DUV without correction, the coupling efficiency collapses to 
0.10 (74.6\% relative drop).
With \name-based ILT (SOCS), we successfully recover the coupling efficiency to 0.337.
These results indicate that \name has the capability of boosting the manufacturability of intrinsically process-sensitive 3D PIC building blocks.

\input{figtex/fig_gc3d}

%% file: figtex/fig_device.tex
\begin{figure}
    \centering
    \includegraphics[width=\columnwidth]{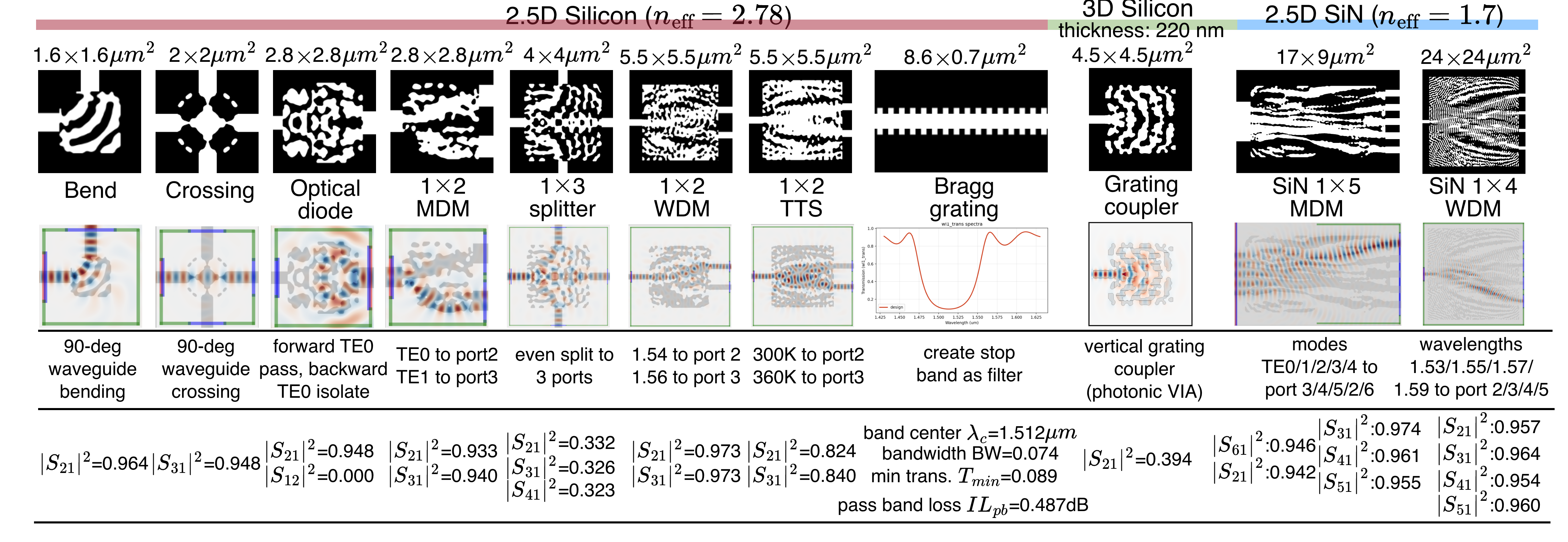}
    \vspace{-13pt}
    \caption{Various Si/SiN inverse-designed, subwavelength photonic device corpus used as benchmarks.
    Design region size is marked on top.
    All 2.5D devices are optimized using the open-source tool MAPS~\cite{ONN_DATE2025_Gu_MAPS} using the effective index (Si: 2.78, SiN: 1.7) with SiO2 as cladding.
    3D grating coupler is inverse designed using Tidy3d.
    }
    \label{fig:device}    
    \vspace{-10pt}
\end{figure}

%% file: tables/tab_ebl_corrected.tex
\begin{table}[t]
\caption{Comparison of four digital twin models trained on three calibration datasets and tested on inverse-designed photonic devices fabricated with \textbf{EBL}, under both \textbf{uncorrected} and \textbf{ILT-corrected} conditions.}
\centering
\vspace{-5pt}
\resizebox{14cm}{!}{
\begin{tabular}{|cc|ccc|ccc|ccc|ccc|}
\hline
\multicolumn{2}{|c|}{\multirow{2}{*}{Training dataset}} & \multicolumn{3}{c|}{CNN (\#params:$\sim 0.4$M)} & \multicolumn{3}{c|}{FNO (\#params:$\sim 0.4$M)} & \multicolumn{3}{c|}{UNet (\#params:$\sim 0.2$M)} & \multicolumn{3}{c|}{PSF (\#params:$5$)} \\ \cline{3-14} 
\multicolumn{2}{|c|}{}                                  & MSE      & PA    & IoU   & MSE      & PA    & IoU   & MSE       & PA    & IoU   & MSE       & PA    & IoU   \\ \hline
\multirow{2}{*}{Normal   patterns} & \multicolumn{1}{|c|}{Test1} & 1.83E-03 & 0.998 & 0.992 & 5.73E-03 & 0.994 & 0.976 & 1.14E-03 & 0.999 & 0.995 & 1.65E-03 & 0.998 & 0.993 \\
                                   & \multicolumn{1}{|c|}{Test2} & 2.45E-03 & 0.998 & 0.990 & 6.56E-03 & 0.993 & 0.972 & 1.34E-03 & 0.999 & 0.994 & 2.02E-03 & 0.998 & 0.991 \\ \hline
\multirow{2}{*}{Random patterns}   & \multicolumn{1}{|c|}{Test1} & 1.51E-03 & 0.998 & 0.994 & 2.55E-03 & 0.997 & 0.989 & 9.84E-04 & 0.999 & 0.996 & 9.64E-04 & 0.999 & 0.996 \\
                                   & \multicolumn{1}{|c|}{Test2} & 2.22E-03 & 0.998 & 0.991 & 3.15E-03 & 0.997 & 0.987 & 1.27E-03 & 0.999 & 0.995 & 1.37E-03 & 0.999 & 0.994 \\ \hline
\multirow{2}{*}{Photonic devices}  & \multicolumn{1}{|c|}{Test1} & 2.23E-03 & 0.998 & 0.991 & 1.90E-03 & 0.998 & 0.992 & 9.99E-04 & 0.999 & 0.996 & 9.82E-04 & 0.999 & 0.996 \\
                                   & \multicolumn{1}{|c|}{Test2} & 2.81E-03 & 0.997 & 0.988 & 2.55E-03 & 0.997 & 0.989 & 1.33E-03 & 0.999 & 0.994 & 1.39E-03 & 0.999 & 0.994 \\ \hline
\end{tabular}
}
\label{tab:ebl_modeling}
\end{table}

%% file: tables/tab_duv_corrected.tex
\begin{table}[t]
\caption{Comparison of four digital twin models trained on three calibration datasets and tested on inverse-designed photonic devices fabricated with \textbf{193 nm DUV}, under both \textbf{uncorrected} and \textbf{ILT-corrected} conditions.}
\vspace{-5pt}
\centering
\resizebox{14cm}{!}{
\begin{tabular}{|cc|ccc|ccc|ccc|ccc|}
\hline
\multicolumn{2}{|c|}{\multirow{2}{*}{Training dataset}} & \multicolumn{3}{c|}{CNN (\#params:$\sim 0.4$M)} & \multicolumn{3}{c|}{FNO (\#params:$\sim 0.4$M)} & \multicolumn{3}{c|}{UNet (\#params:$\sim 0.2$M)} & \multicolumn{3}{c|}{SOCS (\#params:12)} \\ \cline{3-14} 
\multicolumn{2}{|c|}{}                                  & MSE      & PA    & IoU   & MSE      & PA    & IoU   & MSE       & PA    & IoU   & MSE       & PA    & IoU   \\ \hline
\multirow{2}{*}{Normal   patterns} & \multicolumn{1}{|c|}{Test1} & 1.45E-02 & 0.985 & 0.947 & 1.55E-02 & 0.984 & 0.944 & 3.66E-03 & 0.996 & 0.987 & 3.78E-03 & 0.996 & 0.986 \\ 
                                   & \multicolumn{1}{|c|}{Test2} & 2.79E-02 & 0.972 & 0.905 & 2.80E-02 & 0.972 & 0.905 & 6.35E-03 & 0.994 & 0.978 & 7.03E-03 & 0.993 & 0.976 \\ \hline
\multirow{2}{*}{Random patterns}   & \multicolumn{1}{|c|}{Test1} & 2.12E-02 & 0.979 & 0.923 & 1.63E-02 & 0.984 & 0.943 & 3.99E-03 & 0.996 & 0.986 & 3.27E-03 & 0.997 & 0.988 \\
                                   & \multicolumn{1}{|c|}{Test2} & 3.60E-02 & 0.964 & 0.877 & 4.24E-02 & 0.958 & 0.868 & 7.99E-03 & 0.992 & 0.972 & 7.03E-03 & 0.993 & 0.976 \\ \hline
\multirow{2}{*}{Photonic devices}  & \multicolumn{1}{|c|}{Test1} & 1.29E-02 & 0.987 & 0.953 & 1.09E-02 & 0.989 & 0.961 & 3.70E-03 & 0.996 & 0.987 & 4.25E-03 & 0.996 & 0.985 \\
                                   & \multicolumn{1}{|c|}{Test2} & 2.87E-02 & 0.971 & 0.905 & 2.71E-02 & 0.973 & 0.910 & 7.85E-03 & 0.992 & 0.973 & 1.69E-02 & 0.983 & 0.941 \\ \hline
\end{tabular}
}
\label{tab:duv_modeling}
\end{table}

%% file: tables/tab_ebl_ilt.tex
\begin{table}[t]
\caption{Performance and yield improvement from photonic inverse lithography using different fabrication digital twin models under EBL process.}
\vspace{-5pt}
\centering
\resizebox{14cm}{!}{
\begin{tabular}{|c|cc|ccc|cccc|cccc|}
\hline
\multirow{2}{*}{Device}          & \multicolumn{2}{c|}{Device performance (FoM)} & \multicolumn{3}{c|}{w/o ILT}                & \multicolumn{4}{c|}{\name-based ILT (UNet)}                                                                                                & \multicolumn{4}{c|}{\name-based ILT (PSF)}                                                                                         \\ \cline{2-14} 
                                 & before fab.  & after fab.     & L2 error    & $\text{yield}_{80\%}$          & $\text{yield}_{95\%}$          & L2 error                & after fab.        & $\text{yield}_{80\%}$            & $\text{yield}_{95\%}$                           & L2 error                & after fab.                       & $\text{yield}_{80\%}$            & $\text{yield}_{95\%}$                                   \\ \hline
Bending                          & $|S_{21}|^2$: 0.964         & $|S_{21}|^2$: 0.964±0.001  & 1.573E+2 & 100\%                  & 100\%                  &       1.091E+2            & $|S_{21}|^2$: 0.963±0.001  & 100\%                  & 100\%                                   &     1.183E+2             & $|S_{21}|^2$: 0.964±0.010   & 100\%                  & 100\%                    \\   \hline
Crossing                         & $|S_{31}|^2$: 0.948         & $|S_{31}|^2$: 0.947±0.001  & 1.700E+2 & 100\%                  & 100\%                  &       1.128E+2            & $|S_{31}|^2$: 0.948±0.001  & 100\%                  & 100\%                                   &       1.081E+2            & $|S_{31}|^2$: 0.948±0.001  & 100\%                  & 100\%                     \\    \hline
\multirow{2}{*}{Optical diode}   & $|S_{21}|^2$: 0.948         & $|S_{21}|^2$: 0.939±0.002  & \multirow{2}{*}{2.852E+2} & \multirow{2}{*}{100\%} & \multirow{2}{*}{100\%} & \multirow{2}{*}{1.715E+2} & $|S_{21}|^2$: 0.947±0.002  & \multirow{2}{*}{100\%} & \multirow{2}{*}{100\%}                   & \multirow{2}{*}{1.667E+2} & $|S_{21}|^2$: 0.947±0.001  & \multirow{2}{*}{100\%} & \multirow{2}{*}{100\%}  \\
                                 & $|S_{12}|^2$:   0.000      & $|S_{12}|^2$: 0.000±0.000   & &                        &                        &                   & $|S_{12}|^2$: 0.000±0.000 &                        &                                      &                         & $|S_{12}|^2$: 0.000±0.000 &                        &                           \\   \hline
\multirow{3}{*}{1$\times$3 Splitter} & $|S_{21}|^2$: 0.332        & $|S_{21}|^2$: 0.325±0.001 & \multirow{3}{*}{3.953E+2} & \multirow{3}{*}{100\%} & \multirow{3}{*}{100\%} & \multirow{3}{*}{2.297E+2} & $|S_{21}|^2$: 0.333±0.001 & \multirow{3}{*}{100\%} & \multirow{3}{*}{100\%} & \multirow{3}{*}{2.417E+2} & $|S_{21}|^2$: 0.329±0.001 & \multirow{3}{*}{100\%} & \multirow{3}{*}{100\%}  \\
                                 & $|S_{31}|^2$:   0.326      & $|S_{31}|^2$: 0.328±0.001   & &                        &                        &                   & $|S_{31}|^2$: 0.324±0.001 &                        &                                      &                   & $|S_{31}|^2$: 0.325±0.002 &                        &                                         \\
                                 & $|S_{41}|^2$:   0.323      & $|S_{41}|^2$: 0.324±0.001   & &                        &                        &                   & $|S_{41}|^2$: 0.324±0.001 &                        &                                       &                   & $|S_{41}|^2$: 0.326±0.001 &                        &                                  \\   \hline
\multirow{2}{*}{1$\times$2 MDM mux}             & $|S_{21}|^2$: 0.933         & $|S_{21}|^2$: 0.932±0.001  & \multirow{2}{*}{2.805E+2} & \multirow{2}{*}{100\%} & \multirow{2}{*}{100\%} & \multirow{2}{*}{1.644E+2} & $|S_{21}|^2$: 0.933±0.001  & \multirow{2}{*}{100\%} & \multirow{2}{*}{100\%}             & \multirow{2}{*}{1.683E+2} & $|S_{21}|^2$: 0.932±0.001  & \multirow{2}{*}{100\%} & \multirow{2}{*}{100\%}  \\
                                 & $|S_{31}|^2$: 0.940         & $|S_{31}|^2$: 0.929±0.002  &  &                        &                        &                   & $|S_{31}|^2$: 0.939±0.001  &                        &                                         &                   & $|S_{31}|^2$: 0.938±0.001  &                        &                                       \\  \hline
\multirow{2}{*}{ 1$\times$2 WDM mux}             & $|S_{21}|^2$: 0.973         & $|S_{21}|^2$: 0.940±0.016  & \multirow{2}{*}{5.876E+2} & \multirow{2}{*}{86\%}  & \multirow{2}{*}{18\%}  & \multirow{2}{*}{3.400E+2} & $|S_{21}|^2$: 0.964±0.012  & \multirow{2}{*}{100\%} & \multirow{2}{*}{97\%}               & \multirow{2}{*}{3.663E+2} & $|S_{21}|^2$: 0.965±0.010   & \multirow{2}{*}{100\%} & \multirow{2}{*}{96\%}  \\
                                 & $|S_{31}|^2$: 0.973         & $|S_{31}|^2$: 0.861±0.071  &  &                        &                        &                   & $|S_{31}|^2$: 0.957±0.020  &                        &                                          &                   & $|S_{31}|^2$: 0.957±0.024  &                        &                                        \\  \hline
\multirow{2}{*}{1$\times$2 TTS}             & $|S_{21}|^2$: 0.824         & $|S_{21}|^2$: 0.619±0.149  & \multirow{2}{*}{4.806E+2} & \multirow{2}{*}{49\%}  & \multirow{2}{*}{13\%}  & \multirow{2}{*}{2.659E+2} & $|S_{21}|^2$: 0.801±0.048  & \multirow{2}{*}{98\%}  & \multirow{2}{*}{82\%}               & \multirow{2}{*}{2.645E+2} & $|S_{21}|^2$: 0.803±0.044  & \multirow{2}{*}{98\%}  & \multirow{2}{*}{80\%}   \\
                                 & $|S_{31}|^2$: 0.840         & $|S_{31}|^2$: 0.827±0.011  &   &                        &                        &                   & $|S_{31}|^2$: 0.827±0.017  &                        &                                       &                   & $|S_{31}|^2$: 0.824±0.020  &                        &                              \\   \hline
\multirow{4}{*}{Bragg grating}                            &       $\lambda_c$: 1.512            &        $\lambda_c$: 1.516±1.536e-07      &   \multirow{4}{*}{3.176E+2}   &          \multirow{4}{*}{100\%}              &      \multirow{4}{*}{0\%}                  &      \multirow{4}{*}{2.380E+2}              &          $\lambda_c$: 1.513±3.158e-06        &       \multirow{4}{*}{100\%}                 &            \multirow{4}{*}{100\%}                            &           \multirow{4}{*}{2.374E+2}        &       $\lambda_c$: 1.515±1.080e-06           &     \multirow{4}{*}{100\%}                   &             \multirow{4}{*}{70\%}                      \\

     &   $BW$: 0.074                &          $BW$: 0.073±2.439e-08           &         &               &                        &                   &      $BW$: 0.074±4.5658e-08            &                        &                                        &                   &       $BW$: 0.074±1.769e-08           &                        &                                   \\
     
     &        $T_{\min}$: 0.089           &          $T_{\min}$: 0.010±1.108e-06    &       &                        &                        &                   &       $T_{\min}$: 0.087±2.315e-06           &                        &                                        &                   &         $T_{\min}$: 0.094±1.573e-06         &                        &                                   \\
     
     &         $\mathrm{IL}_{\mathrm{pb}}$: 0.487          &       $\mathrm{IL}_{\mathrm{pb}}$: 0.474±2.462e-06     &         &                        &                        &                   &      $\mathrm{IL}_{\mathrm{pb}}$: 0.494±1.033e-05            &                        &                                        &                   &        $\mathrm{IL}_{\mathrm{pb}}$: 0.486±2.433e-06          &                        &                                   \\\hline

\end{tabular}
}
\label{tab:ebl_ilt}
\end{table}

%% file: tables/tab_duv_ilt.tex
\begin{table}[t]
\caption{Performance and yield improvement from photonic inverse lithography using different fabrication digital twin models under 193 nm DUV process.}
\vspace{-5pt}
\centering
\resizebox{14cm}{!}{
\begin{tabular}{|c|cc|ccc|cccc|cccc|}
\hline
\multirow{2}{*}{Device}          & \multicolumn{2}{c|}{Device performance (FoM)} & \multicolumn{3}{c|}{w/o ILT}                & \multicolumn{4}{c|}{\name-based ILT (UNet)}                                                                                                & \multicolumn{4}{c|}{\name-based ILT (SOCS)}                                                                                         \\ \cline{2-14} 
                                 & before fab.  & after fab.     & L2 error     & $\text{yield}_{80\%}$          & $\text{yield}_{90\%}$          & L2 error                & after fab.        & $\text{yield}_{80\%}$            & $\text{yield}_{90\%}$                       & L2 error                & after fab.                       & $\text{yield}_{80\%}$            & $\text{yield}_{90\%}$                                        \\ \hline
Bending                          & $|S_{21}|^2$: 0.964       & $|S_{21}|^2$: 0.755±0.052   &  7.381E+2  & 51\%                 & 0\%                  &       3.032E+2             & $|S_{21}|^2$: 0.963±0.008      & 100\%                  & 100\%                                   &    2.766E+2               & $|S_{21}|^2$: 0.963±0.004                     & 100\%                  & 100\%                                                   \\ \hline
Crossing                         & $|S_{31}|^2$: 0.948       & $|S_{31}|^2$: 0.718±0.051   & 7.275E+2   & 11\%                 & 0\%                  &       3.142E+2            & $|S_{31}|^2$: 0.944±0.006      & 100\%                  & 100\%                                  &     2.735E+2              & $|S_{31}|^2$: 0.944±0.005                     & 100\%                  & 100\%                                                   \\  \hline
\multirow{2}{*}{Optical diode}   & $|S_{21}|^2$: 0.948       & $|S_{21}|^2$: 0.384±0.011   &  \multirow{2}{*}{1.093E+3}  & \multirow{2}{*}{0\%} & \multirow{2}{*}{0\%} & \multirow[c]{2}{*}{4.384E+2} & $|S_{21}|^2$: 0.866±0.087      & \multirow{2}{*}{88\%}  & \multirow{2}{*}{78\%}   & \multirow{2}{*}{3.963E+2} & $|S_{21}|^2$: 0.886±0.069 & \multirow{2}{*}{90\%}  & \multirow{2}{*}{85\%}                    \\
                                 & $|S_{12}|^2$:   0.000    & $|S_{12}|^2$: 0.122±0.006     & &                      &                      &                   & $|S_{12}|^2$: 0.005±0.009 &                        &                                          &                   & $|S_{12}|^2$: 0.004±0.008                &                        &                                                       \\   \hline
\multirow{3}{*}{1$\times$3 Splitter} & $|S_{21}|^2$: 0.332      & $|S_{21}|^2$: 0.089±0.026   &  \multirow{3}{*}{1.544E+3}  & \multirow{3}{*}{0\%} & \multirow{3}{*}{0\%} & \multirow{3}{*}{6.410E+2} & $|S_{21}|^2$: 0.302±0.009     & \multirow{3}{*}{98\%}         & \multirow{3}{*}{50\%} & \multirow{3}{*}{5.724E+2} & $|S_{21}|^2$: 0.312±0.008                    & \multirow{3}{*}{100\%} & \multirow{3}{*}{90\%}                   \\
                                 & $|S_{31}|^2$:   0.326    & $|S_{31}|^2$: 0.072±0.016   &    &                      &                      &                   & $|S_{31}|^2$: 0.323±0.018     &                        &                                       &                   & $|S_{31}|^2$: 0.323±0.020                    &                        &                                                      \\
                                 & $|S_{41}|^2$:   0.323    & $|S_{41}|^2$: 0.115±0.012   &    &                      &                      &                   & $|S_{41}|^2$: 0.322±0.021     &                        &                                        &                   & $|S_{41}|^2$: 0.318±0.014                    &                        &                                                        \\  \hline
\multirow{2}{*}{1$\times$2 MDM mux}             & $|S_{21}|^2$: 0.933       & $|S_{21}|^2$: 0.696±0.031  &  \multirow{2}{*}{1.004E+3}   & \multirow{2}{*}{0\%} & \multirow{2}{*}{0\%} & \multirow[c]{2}{*}{4.602E+2} & $|S_{21}|^2$: 0.924±0.008      & \multirow{2}{*}{100\%}          & \multirow{2}{*}{100\%} & \multirow{2}{*}{3.709E+2} & $|S_{21}|^2$: 0.928±0.008                     & \multirow{2}{*}{98\%}  & \multirow{2}{*}{98\%}                    \\
                                 & $|S_{31}|^2$: 0.940       & $|S_{31}|^2$: 0.430±0.071  &     &                      &                      &                   & $|S_{31}|^2$: 0.898±0.040      &                        &                                         &                   & $|S_{31}|^2$: 0.901±0.054                     &                        &                                                        \\  \hline
\multirow{2}{*}{1$\times$2 WDM mux}             & $|S_{21}|^2$: 0.973       & $|S_{21}|^2$: 0.215±0.045  &  \multirow{2}{*}{2.093E+3}   & \multirow{2}{*}{0\%} & \multirow{2}{*}{0\%} & \multirow{2}{*}{9.409E+2} & $|S_{21}|^2$: 0.736±0.117      & \multirow{2}{*}{28\%}     & \multirow{2}{*}{0\%} & \multirow{2}{*}{7.792E+2} & $|S_{21}|^2$: 0.838±0.131                     & \multirow{2}{*}{54\%}  & \multirow{2}{*}{40\%}  \\
                                 & $|S_{31}|^2$: 0.973       & $|S_{31}|^2$: 0.131±0.031     &         &             &                      &                   & $|S_{31}|^2$: 0.620±0.308      &                        &                                       &                   & $|S_{31}|^2$: 0.676±0.342                     &                        &                                                        \\  \hline
\multirow{2}{*}{1$\times$2 TTS}             & $|S_{21}|^2$: 0.824       & $|S_{21}|^2$: 0.376±0.014   & \multirow{2}{*}{1.752E+3}  & \multirow{2}{*}{0\%} & \multirow{2}{*}{0\%} & \multirow{2}{*}{8.412E+2} & $|S_{21}|^2$: 0.424±0.179      & \multirow{2}{*}{0\%}      & \multirow{2}{*}{0\%} & \multirow{2}{*}{6.510E+2} & $|S_{21}|^2$: 0.563±0.241                     & \multirow{2}{*}{24\%}  & \multirow{2}{*}{16\%}             \\
                                 & $|S_{31}|^2$: 0.840       & $|S_{31}|^2$: 0.441±0.022   &    &                      &                      &                   & $|S_{31}|^2$: 0.651±0.086      &                        &                                         &                   & $|S_{31}|^2$: 0.548±0.267                     &                        &                                                     \\ \hline 
\multirow{4}{*}{Bragg grating}                            &       $\lambda_c$: 1.512            &        $\lambda_c$: 1.545±9.126e-06    &   \multirow{4}{*}{8.872E+2}     &          \multirow{4}{*}{0\%}              &         \multirow{4}{*}{0\%}               &      \multirow{4}{*}{5.602E+2}             &    $\lambda_c$: 1.513±1.453e-05                &            \multirow{4}{*}{28\%}           &            \multirow{4}{*}{0\%}                            &      \multirow{4}{*}{3.646E+2}             &      $\lambda_c$: 1.513±1.414e-05            &         \multirow{4}{*}{90\%}               &            \multirow{4}{*}{79\%}                       \\

     &   $BW$: 0.074                &          $BW$:  0.0524±5.641e-06    &     &                        &                        &                   &       $BW$: 0.068±3.341e-06          &                         &                                        &                   &      $BW$: 0.073±3.434e-06             &                        &                                   \\
     
     &        $T_{\min}$: 0.089           &          $T_{\min}$: 0.363±1.972e-03   &        &                        &                        &                   &         $T_{\min}$: 0.120±4.652e-04         &                        &                                        &                   &       $T_{\min}$: 0.094±2.438e-04           &                        &                                   \\
     
     &         $\mathrm{IL}_{\mathrm{pb}}$: 0.487          &       $\mathrm{IL}_{\mathrm{pb}}$: 0.456±7.426e-04    &         &                        &                        &                   &        $\mathrm{IL}_{\mathrm{pb}}$: 0.485±8.320e-04       &                        &                                        &                   &         $\mathrm{IL}_{\mathrm{pb}}$: 0.487±3.988e-04         &                        &                                   \\\hline

\end{tabular}
}
\label{tab:duv_ilt}
\end{table}

%% file: figtex/fig_wdm_plot.tex
\begin{figure}
    \centering
    \includegraphics[width=0.97\columnwidth]{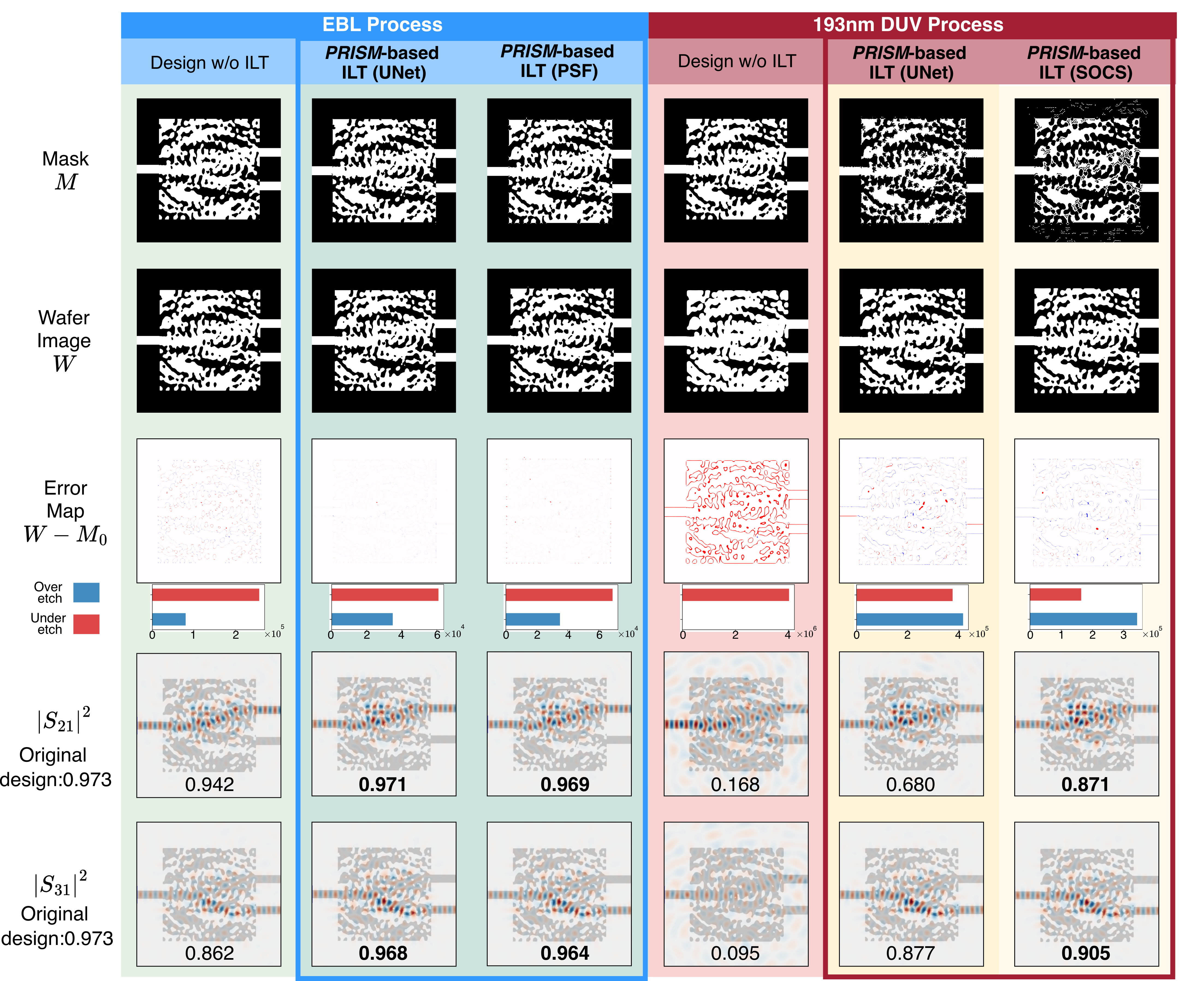}
    \caption{Nominal-case comparison of PRISM-based ILT for an inverse-designed 1$\times$2 Si WDM mux under EBL and 193nm-DUV processes. We show the mask, the wafer image generated by our virtual fab, and the error map, together with the resulting performance metrics ($|S_{21}|^2$ and $|S_{31}|^2$). The inverse-designed device exhibits a substantially larger performance degradation under the DUV process than under EBL. For DUV, the physics-based SOCS lithography model yields more faithful wafer patterns and better recovered device performance than the learned U-Net surrogate. In contrast, for EBL, the learned U-Net and the analytical PSF model achieve comparable pattern fidelity and FoM recovery.}

    \label{fig:wdm_plot}    
\end{figure}

%% file: figtex/fig_grating.tex
\begin{figure*}
    \centering
    \includegraphics[width=0.9\columnwidth]{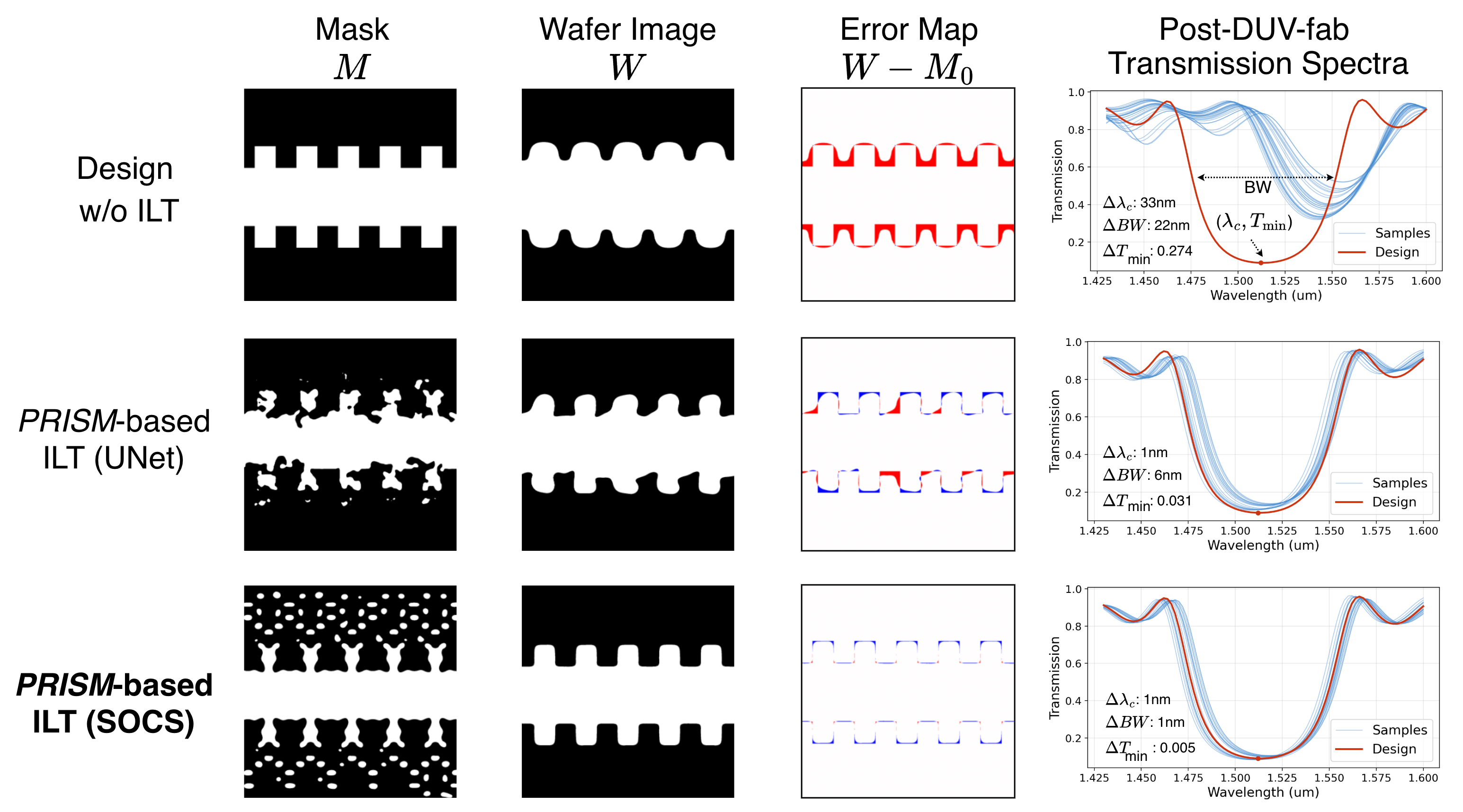}
    \vspace{-10pt}
    \caption{Qualitative and spectra comparison of the uncorrected design and PRISM-based ILT results on DUV-fabricated Si Bragg grating. 
We show the masks, corresponding post-fabrication wafer images, and edge error maps for the nominal design, as well as \name-based ILT (U-Net) and \name-based ILT (SOCS), together with the post-fabrication transmission spectra under sampled process variations (blue) overlaid with the nominal design response (red), illustrating the impact of ILT on pattern fidelity and spectral robustness.
}
    \label{fig:grating_plot}
\end{figure*}

%% file: figtex/fig_yield_optimization.tex
\begin{figure}
    \centering
    \includegraphics[width=1\columnwidth]{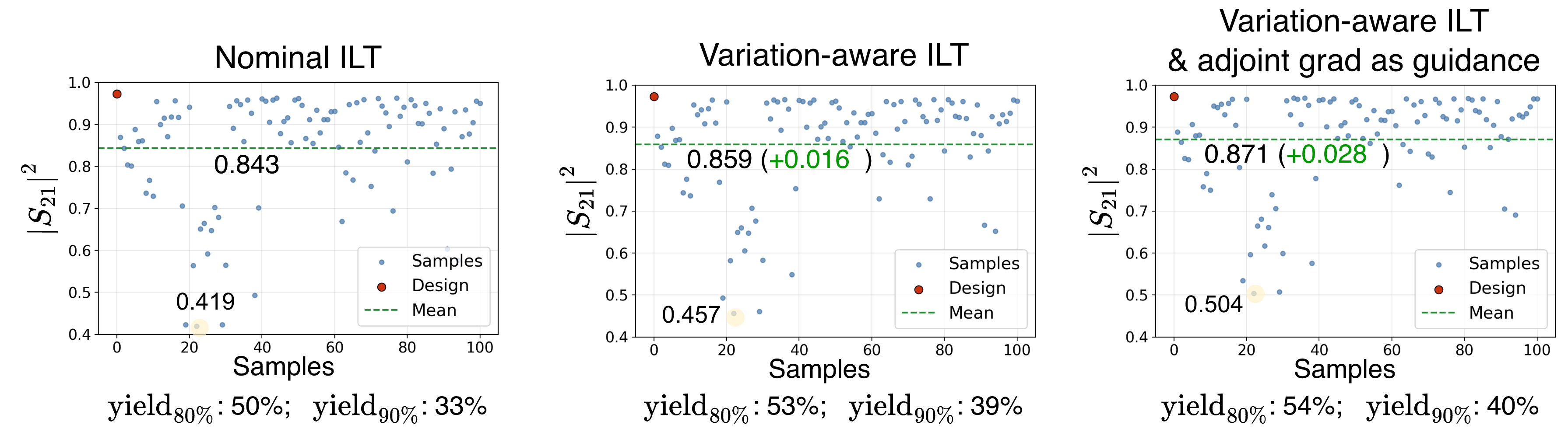}
    \vspace{-13pt}
    \caption{Comparison of the $|S_{21}|^2$ distributions for a $1\times2$ WDM mux after ILT with progressively added variation-aware optimization and adjoint gradient-based guidance.}
    \label{fig:AdjointYield}  
    \vspace{-5pt}
\end{figure}

%% file: figtex/fig_ilt_mask.tex
\begin{figure}
    \centering
    \includegraphics[width=1\columnwidth]{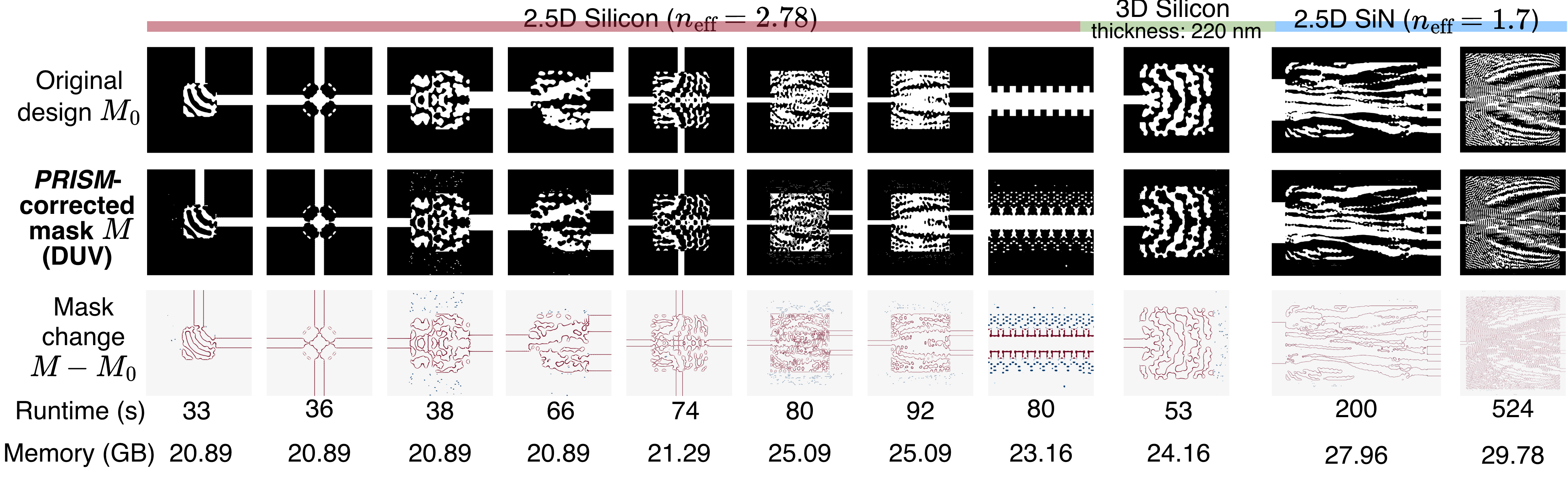}
    \caption{193 nm DUV mask correction results using \name-based ILT (SOCS): for each device, the original design mask $M_0$, the \name-corrected mask, and the change on masks are shown.}
    \label{fig:ilt_mask}    
    \vspace{-8pt}
\end{figure}

%% file: tables/tab_device_area.tex
\begin{table}[tb]

\caption{Performance robustness and ILT effectiveness evaluated on an inverse-designed $1\times2$ WDM mux with various device footprints and minimum feature sizes (MFS).
\centering
}
\resizebox{14cm}{!}{
\begin{tabular}{|cc|cccc|cccc|cccc|}
\hline
\multicolumn{2}{|c|}{1$\times$2 WDM design region size}                                  & \multicolumn{4}{c|}{$5\times5 \mu m^2$}         & \multicolumn{4}{c|}{$6.5\times6.5 \mu m^2$}     & \multicolumn{4}{c|}{$8\times8 \mu m^2$}         \\ \hline
\multicolumn{2}{|c|}{Controlled MFS (nm)}                                  & 50 & 100 & 200 & 300 & 50 & 100 & 200 & 300 & 50 & 100 & 200 & 300 \\ \hline
\multicolumn{1}{|c|}{\multirow{2}{*}{Ideal design mask}} & $|S_{21}|^2$ & 0.957  & 0.968   & 0.951   & 0.929   & 0.983  & 0.980    & 0.980    & 0.970    & 0.994  & 0.996   & 0.997   & 0.988   \\ 
\multicolumn{1}{|c|}{}                                    & $|S_{31}|^2$ & 0.915  & 0.946   & 0.923   & 0.926   & 0.965  & 0.969   & 0.969   & 0.973   & 0.992  & 0.995   & 0.994   & 0.993   \\ \hline
\multicolumn{1}{|c|}{\multirow{2}{*}{Post 193nm-DUV fab}}  & $|S_{21}|^2$ & 0.313  & 0.167   & 0.221   & 0.399   & 0.068  & 0.119   & 0.144   & 0.082   & 0.305  & 0.322   & 0.263   & 0.227   \\ 
\multicolumn{1}{|c|}{}                                    & $|S_{31}|^2$ & 0.036  & 0.031   & 0.110    & 0.094   & 0.193  & 0.141   & 0.278   & 0.081   & 0.007  & 0.005   & 0.013   & 0.022   \\ \hline
\multicolumn{1}{|c|}{\multirow{2}{*}{\name-based ILT (SOCS)}}  & $|S_{21}|^2$ & 0.743  & 0.856   & 0.785   & 0.801   & 0.864  & 0.772   & 0.823   & 0.902   & 0.708  & 0.745   & 0.904   & 0.907   \\ 
\multicolumn{1}{|c|}{}                                    & $|S_{31}|^2$ & 0.770   & 0.536   & 0.732   & 0.567   & 0.812  & 0.631   & 0.742   & 0.824   & 0.512  & 0.675   & 0.857   & 0.857   \\ \hline
\end{tabular}
}
\label{tab:mfs}
\end{table}

%% file: tables/tab_sin.tex
\begin{table}[tb]
\centering
\caption{Performance comparison of SiN $1\times5$ MDM mux and $1\times4$ WDM mux fabricated under EBL and 193nm-DUV processes (nominal case). 
EBL achieves high pattern fidelity and preserves the designed performance, whereas DUV introduces much larger pattern distortions, leading to a pronounced performance degradation.
\name-based ILT can resume the post-fab device performance for both EBL and DUV.}
\vspace{-5pt}
\resizebox{13cm}{!}{
\begin{tabular}{|cc|ccccc|cccc|}

\hline
\multicolumn{2}{|c|}{Device}                                        & \multicolumn{5}{c|}{1$\times$5 MDM mux}   & \multicolumn{4}{c|}{1$\times$4 WDM mux} \\ \hline
\multicolumn{2}{|c|}{Transmission}                                       & TE0   & TE1   & TE2   & TE3   & TE4   & 1530nm    & 1550nm    & 1570nm   & 1590nm   \\ \hline
\multicolumn{2}{|c|}{Ideal design mask}                                   & 0.974 & 0.961 & 0.955 & 0.942 & 0.946 & 0.957   & 0.964   & 0.952  & 0.960   \\ \hline
\multicolumn{1}{|c|}{\multirow{2}{*}{EBL}} & Design w/o ILT         & 0.974 & 0.961 & 0.955 & 0.942 & 0.945 & 0.952   & 0.953   & 0.940   & 0.958  \\
\multicolumn{1}{|c|}{}                     & \name-based ILT (UNet) & 0.974 & 0.961 & 0.955 & 0.942 & 0.946 & 0.954   & 0.964   & 0.950   & 0.959  \\ \hline
\multicolumn{1}{|c|}{\multirow{2}{*}{DUV}} & Design w/o ILT         & 0.936 & 0.925 & 0.924 & 0.901 & 0.912 & 0.057   & 0.016   & 0.021  & 0.036  \\
\multicolumn{1}{|c|}{}                     & \name-based ILT (SOCS) & 0.972 & 0.952 & 0.953 & 0.940  & 0.942 & 0.915   & 0.843   & 0.867  & 0.916  \\ \hline
\end{tabular}
}
\label{tab:sin_ilt}
\vspace{-5pt}
\end{table}

%% file: figtex/fig_sin.tex
\begin{figure*}
    \centering
    \subfloat[]{\includegraphics[width=0.99\linewidth]{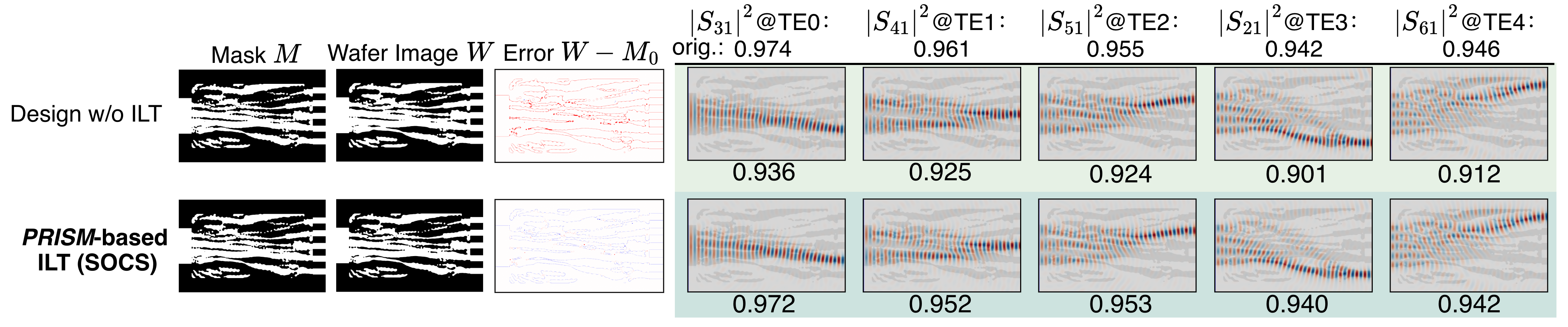}
    \label{fig:SiN_MDM}
    }
    \vspace{-10pt}
    \subfloat[]{\includegraphics[width=0.99\linewidth]{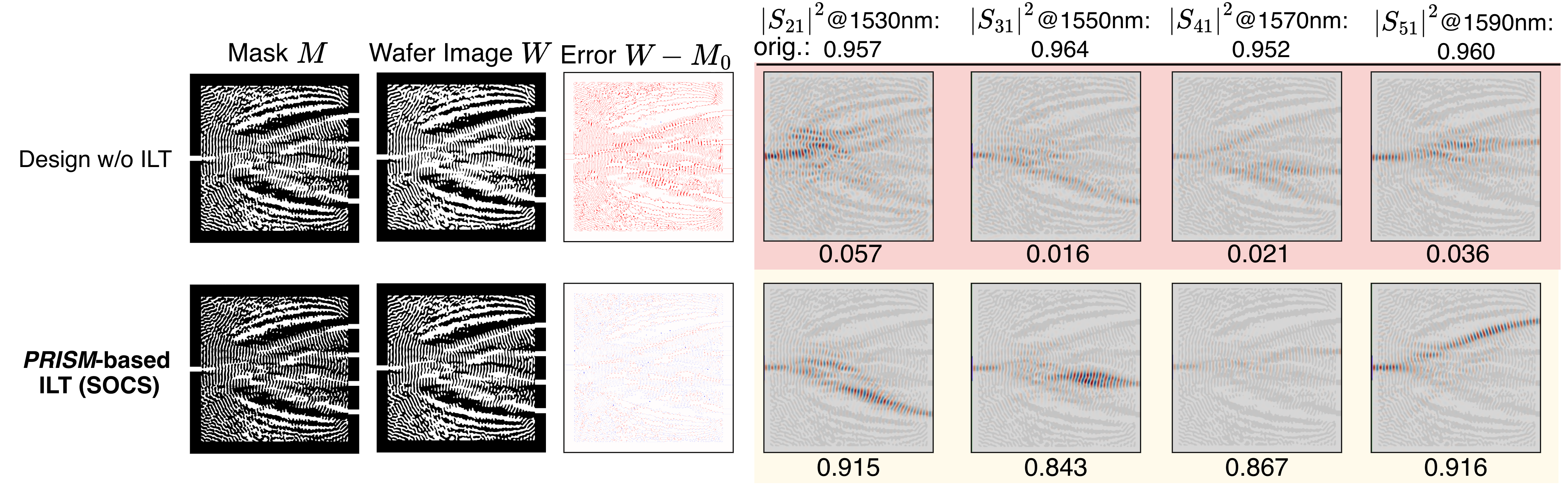}
    \label{fig:SiN_WDM}
    }
    \vspace{-10pt}
    \caption{
    Nominal-case 193nm-DUV virtual-fabrication results for two SiN devices: (a) a $1\times5$ MDM ($17~\mu\mathrm{m}\times9~\mu\mathrm{m}$) and (b) a $1\times4$ WDM ($24~\mu\mathrm{m}\times24~\mu\mathrm{m}$). For each device, we visualize the input mask $M$, the virtual-fabricated wafer image $W$, and the error map $W-M$, along with the corresponding S-parameters. 
    Owing to the lower index contrast (SiN/SiO2) and high-order TE mode profiles, the SiN MDM is relatively insensitive to nominal DUV process distortions and maintains high performance even without correction. 
    In contrast, the grating-based WDM is highly process-sensitive and fails without correction, whereas \name-based ILT (SOCS) significantly reduces pattern errors and recovers the desired spectral response.}
    \label{fig:SiN}
\end{figure*}

%% file: figtex/fig_gc3d.tex
\begin{figure}[tb]
    \centering
    \includegraphics[width=0.95\columnwidth]{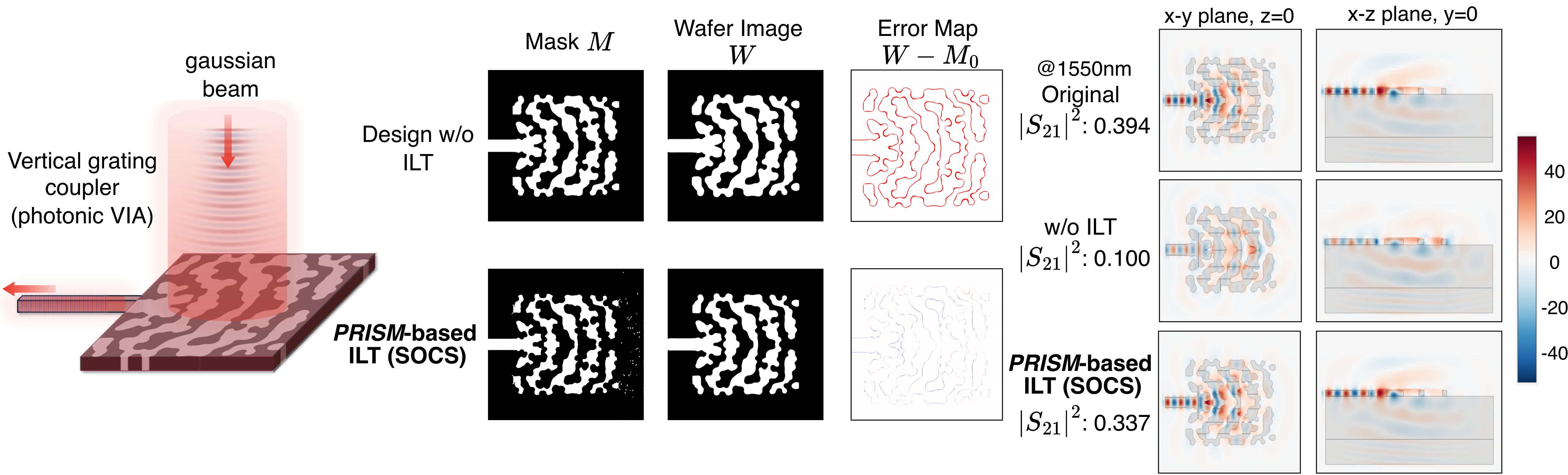}
    \caption{193nm DUV virtual fabrication results for an inverse-designed 3D Si vertical grating coupler optimized with Tidy3D. We compare the uncorrected design and the \name-based ILT (SOCS) mask, together with their corresponding virtual DUV-fabricated wafer images $W$, error maps $W-M_0$, and the 3D FDTD-simulated field and coupling transmission.}
    \label{fig:gc3d}    
\end{figure}

%% file: doc/5_conclu.tex
\section{Conclusion}
\label{sec:Conclusion}
We presented \name, a photonics-informed inverse lithography framework for curvilinear, inverse-designed photonic components that treats \emph{device performance}, not mere geometric fidelity, as the primary objective, enabling a dramatic yield boost (from near-zero to near-unity in representative cases) under realistic fabrication conditions across both EBL and 193 nm DUV processes.
Our study delivers three advances that shift ILT from \emph{rectilinear physics-agonstic geometry repair} to \emph{curvilinear functionality-preserving yield optimization} in inverse-designed photonics.
(i) We elevate \emph{gradient reliability} to a first-class requirement and show that physics-grounded digital twins can provide stable first-order guidance for ILT on the out-of-distribution masks generated during iterative correction.
(ii) We demonstrate that this capability remains practical under realistic, scarce-data conditions by synthesizing compact, physics-guided calibration motifs that efficiently capture non-ideal fabrication behaviors at low fabrication, metrology, and labeling cost, lowering the barrier for both designers and foundries.
(iii) We further inject physics priors via adjoint sensitivity weighted, variation-aware ILT objectives, explicitly optimizing yield and worst-case performance rather than relying on physics-agnostic, geometry-only correction.
Overall, \name advances photonic design-technology co-optimization by elevating photonics ILT to a first-class EPDA stage and bridging the simulation-to-fabrication gap for inverse-designed, process-sensitive photonic components.
By enabling scalable, high-yield photonic manufacturing under both EBL and 193 nm DUV, \name accelerates the lab-to-fab translation of inverse-designed photonic devices and expedites the path toward large-scale photonic integration.

%% file: doc/6_acknow.tex
\begin{acks}
  This work was in partial support by the Empire State Development's Division of Science, Technology and Innovation (NYSTAR) Focus Center at RPI, C210117. The authors would like to thank the support of the SMART fellowship program.
\end{acks}